\documentclass[graybox, nosecnum]{svmult}

\usepackage{mathptmx}       
\usepackage{helvet}         
\usepackage{courier}        
\usepackage{type1cm}        
\usepackage{makeidx}         
\usepackage{graphicx}        
\usepackage{multicol}        
\usepackage[bottom]{footmisc}
\usepackage{hyperref}        
\usepackage{soul}            
\usepackage{amsmath}         
\usepackage{subfig}          
\hypersetup{colorlinks=true,urlcolor=blue}
\usepackage[square,numbers]{natbib}
\makeindex             

\begin{document}
\title*{Basics of Fourier Analysis for High-Energy Astronomy}
\author{Tomaso M. Belloni \thanks{corresponding author} and Dipankar Bhattacharya}
\institute{Tomaso M. Belloni \at INAF - Osservatorio Astronomico di Brera, via E. Bianchi 46, I-23807 Merate, Italy, \email{tomaso.belloni@inaf.it}
\and Dipankar Bhattacharya \at Inter-University Centre for Astronomy and Astrophysics, Post Box No. 4, Ganeshkhind, Pune-411007, India,  \email{dipankar@iucaa.in}}
%
%
\maketitle
\abstract{The analysis of time variability, whether fast variations on time scales well below the second or slow changes over years, is becoming more and more important in high-energy astronomy. Many sophisticated tools are available for data analysis and complex practical aspects are described in technical papers. Here, we present the basic concepts upon which all these techniques are based. It is intended as a condensed primer of Fourier analysis, dealing with fundamental aspects that can be examined in detailed elsewhere. It is not intended to be a presentation of detailed Fourier tools for data analysis, but the reader will find the theoretical basis to understand available analysis techniques.}
\section{Keywords} 
Fourier theory, time-series analysis, high-energy astronomy
\tableofcontents{}
\newpage
\markboth{Basics of Fourier Analysis}{Tomaso Belloni and Dipankar Bhattacharya}

\section{Fourier 101}

Two centuries ago, in 1822 the French mathematician Jean-Baptiste Joseph Fourier publishes his book on the theory of heat, where he claims that all functions, including those containing discontinuities, can be expresses as a series of sinusoids. This is a very important fact, as it allows to deal with functions that otherwise would be intractable. However, it is not formally correct, as the function must satisfy a set of conditions, formulated by Dirichlet a few years later (see below).
The application of Fourier theory nowadays is so widespread in science that it can be considered one of the main scientific discoveries in history. It would be impossible to list all fields where its application is essential, from telecommunication to image analysis. Here we will limit ourselves to what is needed for the analysis of {\it time series}. By time series we mean any one-dimensional signal as a function of {\it time}. It is not important what this signal measures, it can be the varying flux of a cosmic X-ray source, the Dollar-to-Euro exchange rate, the number of points scored by a basketball team. In principle, it is not even important that the independent variable is time, as long as we are dealing with a one-dimensional measurement, but it is simple to remain closer to what is needed for timing analysis.
Specifically, this section of the Handbook is devoted to timing analysis of high-energy astronomical data, but what we describe here is more general.

In this chapter, we will present the basic aspects of Fourier theory, which are at the base of all methods of analysis of time variability that are described in other chapters. We will not describe techniques that can be applied to data analysis, but the principles that these techniques are based upon and that should be known when they are applied. We will by no means be exhaustive, but cover the main properties of Fourier analysis. For more details connected to this topic there are excellent books \citep{butz,jenkinswatts,weaver}.
\subsection{Fourier series}
\label{subsec:Timing:Basic:FourierSeries}

The aim of Fourier analysis is to express any function as a sum of different sines and cosines , characterised by an angular frequency $\omega$ or a corresponding time period $P=2\pi/\omega$.  Therefore, it is easier to visualise the decomposition of an arbitrary periodic function into its sine and cosine components.  Such a decomposition is known as a Fourier series.

A periodic function is defined by its shape within a basic period $P_0$, which is then repeated ad infinitum at the interval of $P_0$.  If we define $\omega_0=2\pi/P_0$, then $\sin\omega_0 t$ and $\cos\omega_0 t$ share the same repetitive property and the sum 
\begin{equation}
    a\sin\omega_0 t + b\cos\omega_0 t  \label{eq:fundamental_sinusoids}
\end{equation}
can represent an infinite (but not exhaustive) variety of periodic functions with period $P_0$ by arbitrarily adjusting the values of $a$ and $b$.  One may then note that sines and cosines with frequencies that are integral multiples of $\omega_0$ also repeat with this period, albeit repeating also within the period.  When all such functions are put together in the decomposition, namely
\begin{equation}
    f(t) = \frac{a(0)}{2}+\sum_{n=1}^\infty [ a(n)\sin n\omega_0 t + b(n)\cos n\omega_0 t] \label{eq:fourierseries_sinusoids}
\end{equation}
then it can represent practically any periodic function with period $P_0$, with suitable adjustment of the coefficients $a(n)$ and $b(n)$, which are known as the Fourier coefficients.

The conditions under which the Fourier decomposition is valid are that $f(t)$ has only a finite number of finite discontinuities and only a finite number of extreme values within a period.  These are known as Dirichlet conditions and the functions obeying them are called piece-wise regular. Since sines and cosines are continuous functions, at the location of a discontinuity in $f(t)$, its Fourier series representation (eqn.~\ref{eq:fourierseries_sinusoids}) evaluates to a definite value, which is the average of the left and right limits of the original function.

The Fourier coefficients $a(n)$ and $b(n)$ can be determined by performing the following integrations:
\begin{equation}
    a(n) = \frac{2}{P_0} \int_0^{P_0} f(t)\sin n\omega_0 t \, dt \;\; ; \;\;\; b(n) = \frac{2}{P_0} \int_0^{P_0} f(t)\cos n\omega_0 t \, dt, \;\; n=0,1,2...  \label{eq:fourierseries_coefficients}
\end{equation}

Alternatively, the Fourier decomposition may also be expressed in an equivalent exponential form:
\begin{equation}
    f(t) = \frac{1}{P_0}\sum_{n=-\infty}^{\infty} c(n) e^{i n\omega_0 t}; \;\;\mbox{\rm where}\;\; c(n)=\int_{0}^{P_0} f(t) e^{-in\omega_0 t} dt \label{eq:fourierseries_exponential}
\end{equation}

In the above expansions, the $n=0$ term is often called the {\em constant} or the {\em D.C.} (Direct Current) component, the $n=1$ term the {\em fundamental} and the terms with $n>1$ the {\em harmonics}.

\begin{figure}
    \centering
    	\includegraphics[width=\columnwidth]{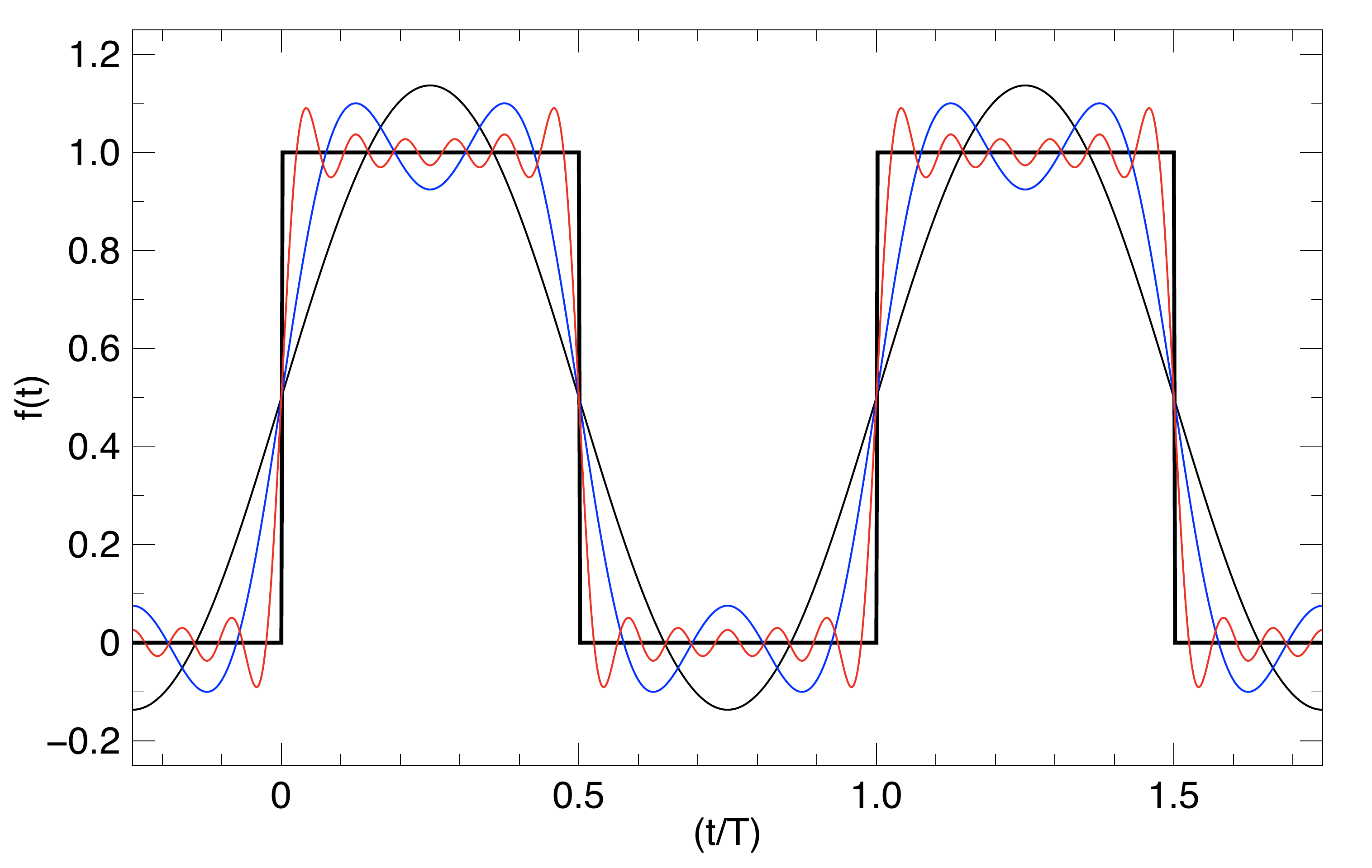}
    \caption{Example of a Fourier series decomposition. The original time domain function $f(t)$, shown in the thick black line, is a square wave of period $T$, which equals unity from $t=0$ to $0.5T$ and zero from $0.5T$ to $T$.   Two periods of the function are plotted. Thin lines show the Fourier sum with different number of terms: (i) black: DC$+$Fundamental, (ii) blue: sum up to 3rd harmonic, (iii) red: sum up to 11th harmonic. The gradual improvement in the approximation of the function with increasing number of terms in Fourier series is evident. At the point of discontinuity, all the reconstructions pass through the average of the left and right limits of the original function and the transition gets progressively sharper with larger number of terms.}
    \label{fig:square_wave_fourier_decomposition}
\end{figure}

Such a decomposition is possible for any periodic function because the sine and cosine functions, or the exponentials, involved in the above expressions form a complete orthogonal basis set.  Further, considering $t$ on the entire real axis $[-\infty,\infty]$ we note that sines and cosines of arbitrary frequency $\omega$ form also a complete, orthogonal set and so do the corresponding exponentials.  It should therefore be possible to express any well-behaved function $f(t)$ in terms of these bases.
\subsection{Continuous Fourier transform}
Now we are ready to extend the decomposition to any function $f(t)$, expanding the integral over the entire real axis.
We define the {\it Fourier transform} (FT) of a function $f(t)$ as 

\begin{equation}
F(\omega) = \int\limits_{-\infty}^{+\infty} f(t) e^{-i\omega t} dt
\label{eq:cont_ft}
\end{equation}

where $\omega = 2\pi\nu$ is the {\it angular frequency} (in radians) and $\nu$ is the frequency (in Hz).
Equation \ref{eq:cont_ft} is a linear transformation and no information is lost. The representations of a function in time and frequency domains are equivalent. The original $f(t)$ can be recovered by applying the {\it inverse Fourier transform}

\begin{equation}
f(t) = {1\over 2\pi} \int\limits_{-\infty}^{+\infty} F(\omega) e^{i\omega t} d\omega
\label{eq:cont_inv_ft}
\end{equation}

Since the expression $e^{i\omega t}$ is a sinusoid, this is equivalent to the decomposition of the original signal into a sum of sinusoids. If the original signal is itself a sinusoid with frequency $\omega_0$, it is easy to see that its transform is $\delta(\omega-\omega_0)$.

The transform has a number of interesting properties. It is linear, not necessarily a real function, and its amplitude is invariant to time shift (but not its phase). Moreover, what is shown in Tab. \ref{tab:FTproperties} applies (throughout the chapter we will use lowercase letters for functions in the time domain and uppercase for frequency domain. The complex conjugate of a function $f$ will be indicated as $f^*$).

\begin{table}
	\centering
	\caption{Basic properties of the continuous Fourier transform}
	\label{tab:FTproperties}
	\begin{tabular}{|c|c|} 
		\hline
		f(t)                         & F($\omega$))\\
		\hline
        Real                         & H(-f) = H$^*$($\omega$) \\
        Even                         & Even \\
        Odd                          & Odd \\
        Real \& Even                 & Real \& Even  \\
        Real \& Odd                  & Imaginary \& Odd \\
		\hline
	\end{tabular}
\end{table}

Unless the original function is even, an unlikely situation in the case of a time series, its Fourier transform is complex: each sinusoid at angular frequency $\omega$ is characterized by its amplitude and its phase. Since the phases are meaningful only in the case of periodic signals, what is usually considered for Fourier analysis is the {\it Power Density Spectrum (PDS)}, defined as the Fourier Transform multiplied by its complex conjugate and therefore the square modulus of the Fourier Transform:

\begin{equation}
P(\omega) = F(\omega)\cdot F^*(\omega) =  | F(\omega) |^2
\label{eq:cont_pds}
\end{equation}

If the original function is real, which is of course usually the case for time series, the PDS is an even function (see Tab. \ref{tab:FTproperties}), so the values at negative frequencies are redundant.
Notice that although the FT is a linear function, the PDS obviously is not. This means that while the FT of the sum of two signals is the sum of the FT of the signals, in the case of the PDS this is not true and there are cross-terms to be considered. More specifically, if our two signals are $f(t$) and $g(t)$, the PDS of the sum of the two is

\begin{equation}
\begin{split}
    P[f(t)+g(t)] & = |F[f(t)+g(t)]|^2 =  F[f(t)+g(t)]\cdot F^*[f(t)+g(t)] = \\
              & = P[f(t)] + P[g(t)] + 2 Re\{F[f(t)]\cdot F[g(t)] \}
    \end{split}
    \label{eq:pds_sum}
\end{equation}

If the two signals are uncorrelated, the cross-term is zero and linearity applies.

The PDS is by definition a real function. We can see some simple example. If $f(t) = a\cdot cos(\omega_0 t)$, its transform is $F(\omega) = a\cdot \delta(\omega-\omega_0)$ (a real function as the original function is Real and Even (see Tab.\ref{tab:FTproperties}) and its PDS is $P(\omega) = a^2 \cdot \delta(\omega-\omega_0)$.
If the original function is a one-sided exponential

\begin{equation}
f(t) = 
\begin{cases}
e^{-\lambda t} & t \geq 0\\
0                    & t < 0
\end{cases}
\label{eq:exponential}
\end{equation}

its tranform is

\begin{equation}
F(\omega) = {1\over i\omega + \lambda}
\label{eq:ft_exponential}
\end{equation}

and its PDS is

\begin{equation}
P(\omega) = {1\over \omega^2 + \lambda^2}
\label{eq:pds_exponential}
\end{equation}

since we are interested here in functions of time, it is interesting to remark,  that the human ear works essentially in the same way: the hairs on the organ of Corti in the inner ear vibrate and record the intensity of the incoming sound, but are not sensitive to the phase. Effectively, a PDS is produced to be transmitted to the brain.

The {\it autocorrelation (ACF)} of a function $f(t)$, which we will examine in more detail later, is defined as

\begin{equation}
A(t) = \int\limits_{-\infty}^{+\infty} f(\tau) f(t+\tau) d\tau \Longleftrightarrow F(f) F^*(f) \equiv |F(f)|^2
\label{eq:acf}
\end{equation}

Equation \ref{eq:acf} shows that the autocorrelation of a function is the Fourier transform of its PDS (here $^*$ means complex conjugate and $\Longleftrightarrow$ means ``is the Fourier transform of"). Since the PDS of a real function is real and even, Tab. \ref{tab:FTproperties} tells us that the ACF is also real and even.
From Eqn. \ref{eq:acf} it is simple to derive {\it Parseval's theorem} (simply setting t=0)

\begin{equation}
\int\limits_{-\infty}^{+\infty} |f(t)|^2 dt = {1\over 2\pi} \int\limits_{-\infty}^{+\infty} |f(\omega)|^2 d\omega
\label{eq:cont_parseval}
\end{equation}

This is important: the ACF of a function and its PDS are equivalent and contain the same amount of information.

\subsection{Discrete Fourier transform}

What we have shown so far is interesting mathematically, but it is rather abstract as it deals with continuous functions, possibly even in the complex domain, extending from $-\infty$ to $+\infty$. What we have in X-ray astronomy is discrete measurements extending from 0 to T: a time series (commonly called ``light curve") consisting of $N$ measurements $x_k$ taken at equally-spaced times $t_k$ from $0$ to $T$ (we will see later what happens if there are gaps, or the times are not equally spaced). In this case we can move to the {\it discrete} Fourier transform (and its inverse), defined as

\begin{align}
a_j & =            \sum\limits_{k=0}^{N-1} x_k e^{-2\pi ijk/N}&\quad\quad (j=-N/2,...,N/2-1) \label{eq:Timing:Basic:DFT}\\
x_k & = {1\over N} \sum\limits_{k=-N/2}^{N/2-1} a_j e^{2\pi ijk/N}&\quad\quad (k=0,...,N-1)
\end{align}

As in the case of the continuous version, there is no loss of information: N numbers in the signal, N numbers in the FT. The FT is in general a complex quantity even for a real signal, doubling the information per frequency, but the positive and negative values are clearly correlated.

Since the data are equally spaced, the times are $kT/N$ and the frequencies are $j/T$. The time step is $\delta t = T/N$ and the frequency step is $\delta\nu =1/T$. This is a very useful version of the transform as it can be applied to data rather than functions, but this comes with limitations. As the discrete time series has a time step $\delta t$ and a duration $T$, there are limitations to the frequencies that can be examined. 
The lowest frequency is $1/T$, corresponding to a sinusoid with a period equal to the signal duration. The highest frequency that can be sampled, is called {\it Nyquist} frequency (Eqn. \ref{eq:nyquist}) 

\begin{equation}
    \nu_{Nyq} = {1\over 2\delta t} = {1\over 2} {N\over T}.
    \label{eq:nyquist}
\end{equation}

It corresponds to a sinusoid sampled twice per cycle, therefore appearing as up-and-down in the original signal.
Notice that $\delta\nu =1/T$ means that the frequency resolution of your transform is inversely proportional to the duration of the signal and does not depend on the signal sampling.

There is a zero frequency, at which the FT value is simply the sum of the signal values
\begin{equation}
    a_0 = \sum\limits_{k=0}^{N-1} x_k e^{-2\pi i0k/N} = \sum\limits_{k=0}^{N-1} x_k
    \label{eq:zerofrequency}
\end{equation}

\begin{equation}
    P_j = |a_j|^2
    \label{eq:disc_pds}
\end{equation}

Notice that the PDS at negative frequencies is identical to that at positive frequencies, as in the case of its continuous version. 

Parseval's theorem applies also to the discrete case:

\begin{equation}
    \sum\limits_{k=0}^{N-1} |x_k|^2 = {1\over N} \sum\limits_{k=-N/2}^{N/2-1}|a_j|^2
    \label{eq:disc_parseval}
\end{equation}

from which one can see that the variance of the signal is $1/N$ times the sum of the $a_j$ over all indices besides zero (see \cite{vdk88}).
Also the connection with the discrete autocorrelation applies: the PDS is the Fourier transform of the ACF.

\begin{figure}
	\includegraphics[width=\columnwidth]{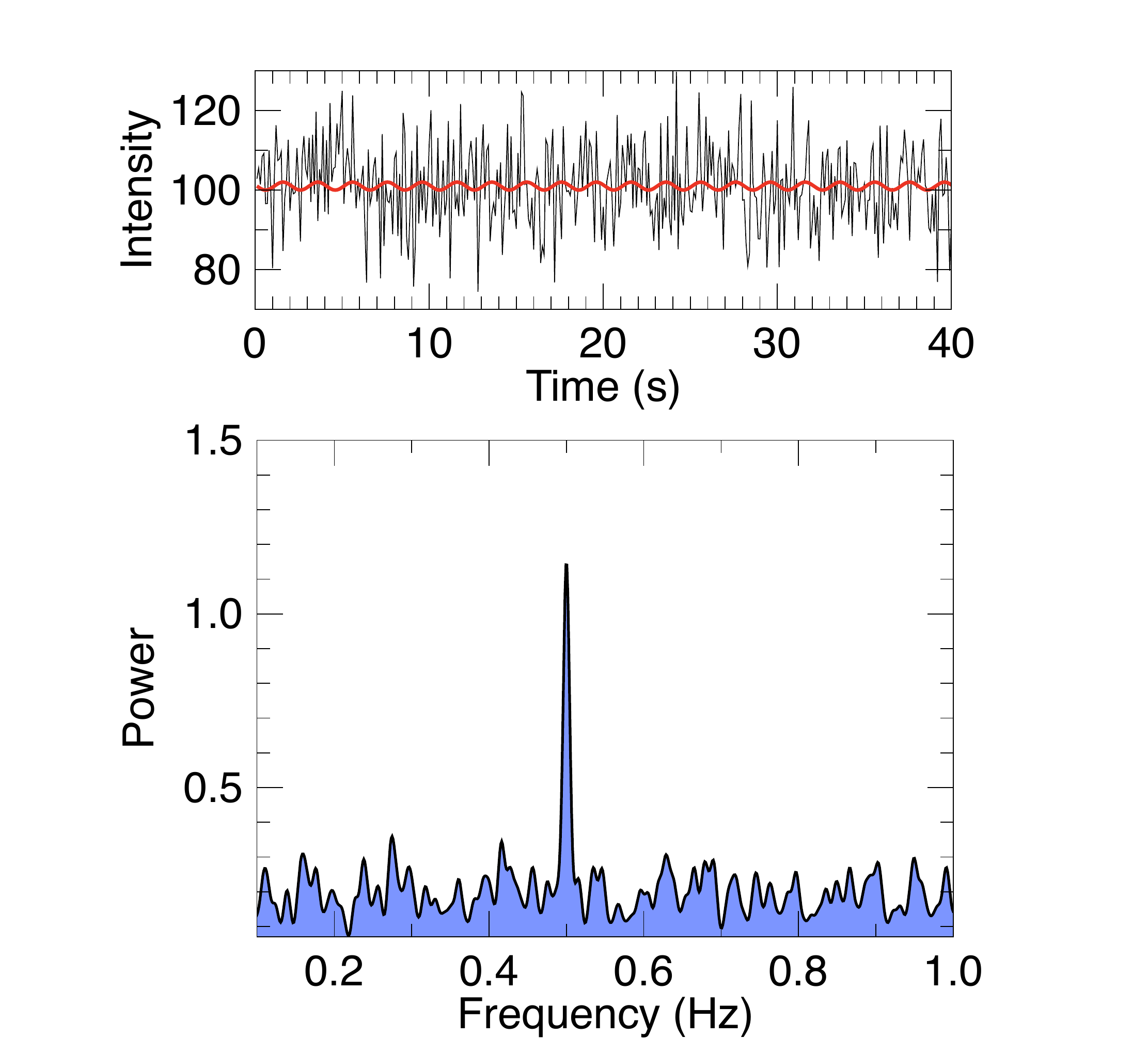}
    \caption{Top panel: the black line shows 40 seconds of a simulated 1000-s time series consisting of a weak sinusoidal modulation (red line) "drowned" into a strong Gaussian noise. Bottom panel: corresponding PDS, zoomed on the relevant frequency range, where the modulation is clearly visible. A weak signal spread in time is collected into a single frequency bin at high significance.
    }
    \label{fig:power_of_PDS}
\end{figure}

Fourier theory is an extremely powerful tool for extracting information from time series. An example can be seen in Fig. \ref{fig:power_of_PDS}. The top panel shows part of a simulated time series (black line) consisting of strong Gaussian noise superimposed on a weak sinusoidal modulation (red line). The modulation is completely invisible by eye, but appears clearly in the PDS (bottom panel). 

\subsection{Windowing and sampling}

We have seen two definitions of Fourier Transform: the continuous FT, which applies to functions over the ($-\infty,\infty$) interval, and the discrete FT, which deals with $N$ sampled data over the $(0,T)$ range. The question now is how to connect the first, which has numerous properties but is not realistic for data analysis, to the second. In order to do this we can apply one of the fundamental properties of the continuous FT: the Fourier Transform of the product of two functions $x(t)$ and $y(t)$ is the convolution of the Fourier transforms of the functions:

\begin{equation}
    F[x \cdot y] = F[x] \otimes F[y] = \int\limits_{-\infty}^{+\infty}F[x(\nu')]F[y(\nu-\nu')] d\nu'
    \label{eq:convolution}
\end{equation}

Our discrete time series $x_k$ vs. $t_k$ can be seen as the product of a continuous function $f(t)$ over ($-\infty,\infty$) and two additional functions: $w(t)$ to limit it to the $(0,T)$ interval and $s(t)$ to sample it at times $t_k$:

\begin{equation}
    x_k = h(t) \cdot w(t) \cdot s(t)
    \label{eq:cont_too_disc}
\end{equation}

\begin{figure}
	\includegraphics[width=\columnwidth]{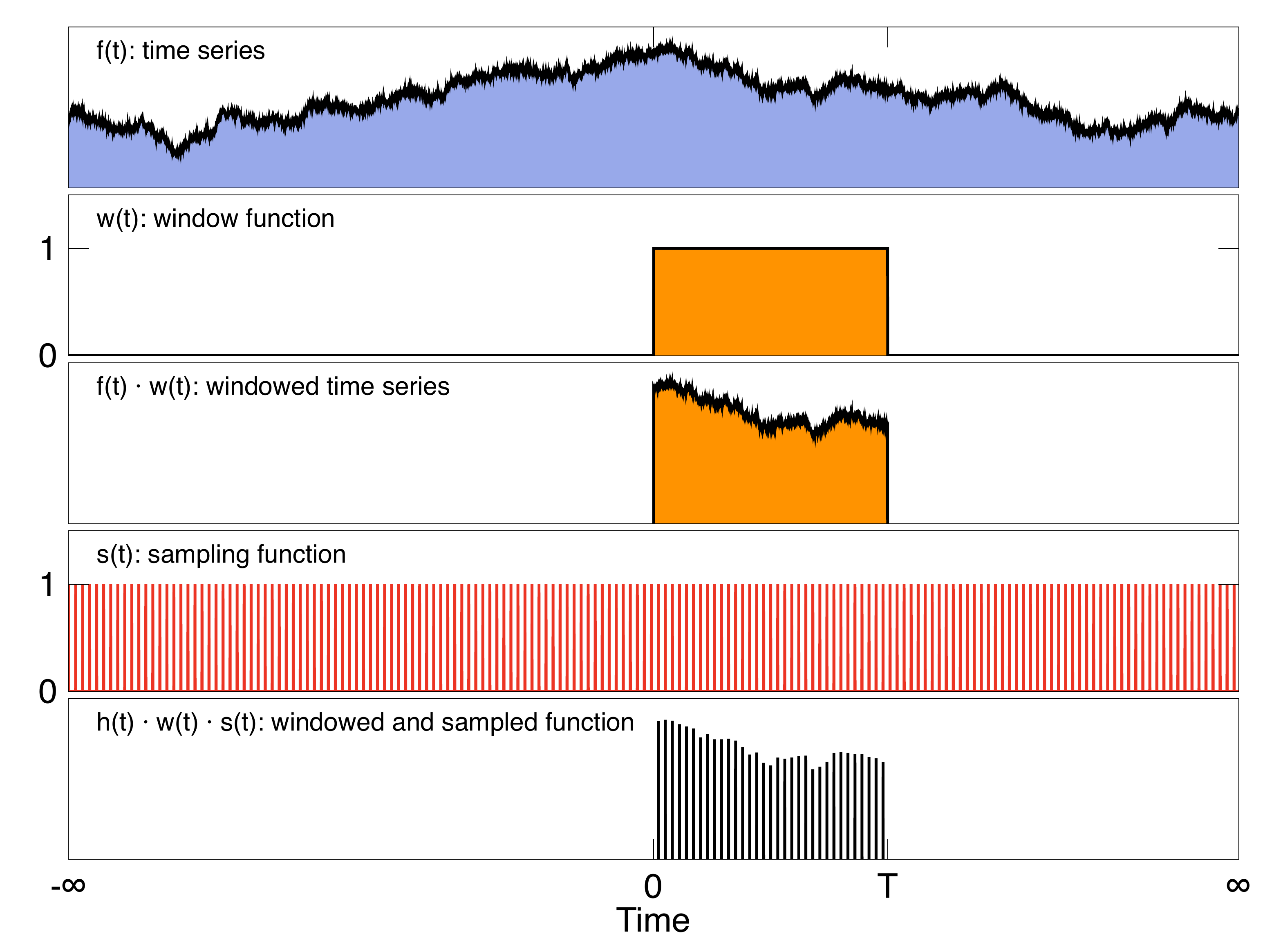}
    \caption{From continuous function to discrete data. The panels from top to bottom : continuous function $f(t)$ over the $(-\infty,\infty)$ interval; boxcar window function $w(t)$; $f(t) \cdot w(t)$ windowed continuous function; $s(t)$ sampling function; final windowed and sampled function, corresponding to the discrete data.
    }
    \label{fig:windowing}
\end{figure}

$w(t)$ is a boxcar window function (more on windows below), which is 1 in the $(0,T)$ interval and zero outside. $s(t)$ is a series of delta functions at $t_k$, spaced by $T/N$ (see above). It is simpler to show this graphically in Fig. \ref{fig:windowing}.

\subsubsection{Windowing effects}
    
\begin{figure}
	\includegraphics[width=\columnwidth]{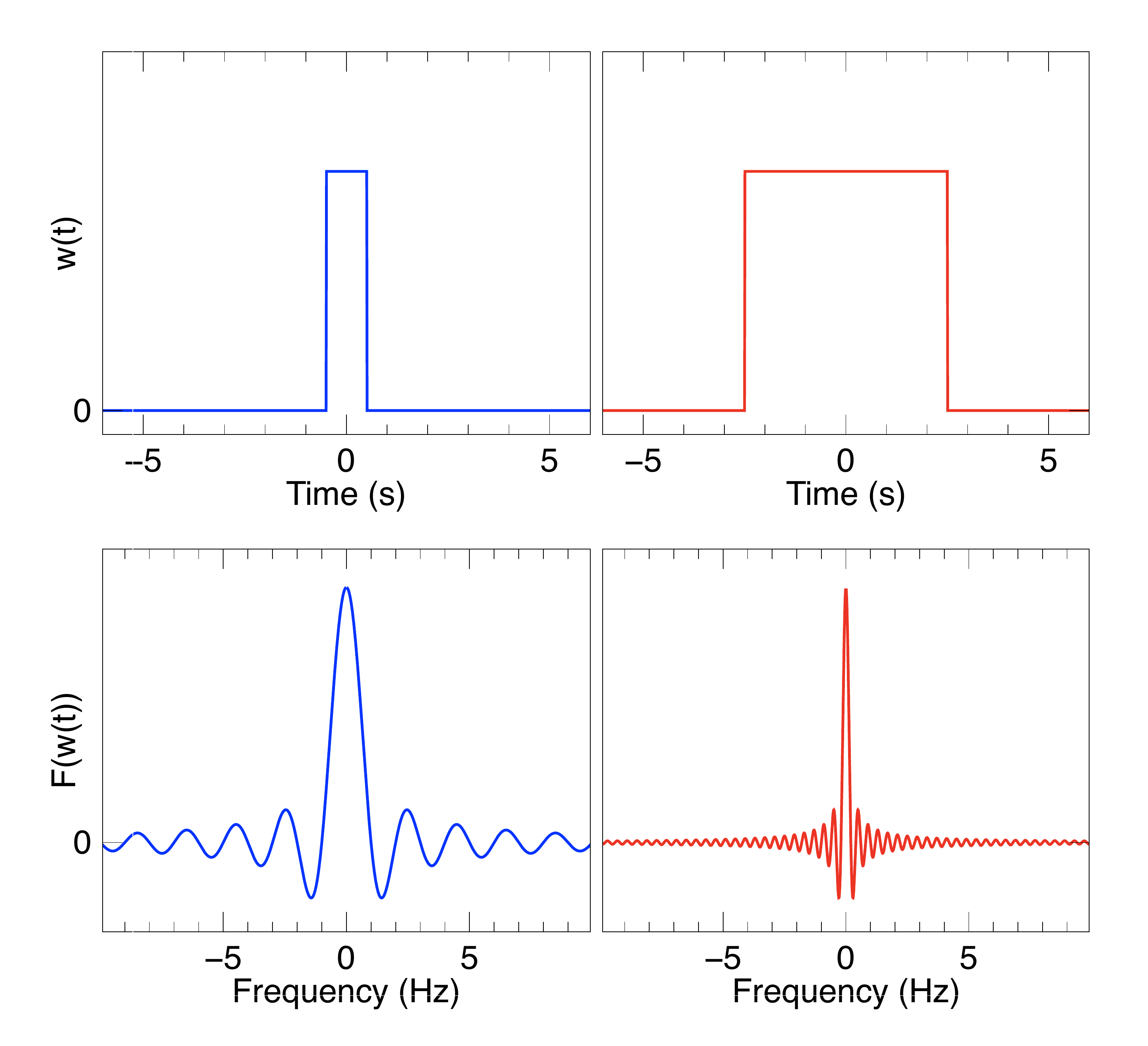}
    \caption{Top panels: two boxcar windows with duration 1 s (blue) and 5 s (red). Bottom panels: their respective FT.
    }
    \label{fig:window_ft}
\end{figure}

To see what the effect of windowing and sampling is let us consider a purely sinusoidal function $f(t) = sin(\omega t)$, whose FT is a delta function at $\omega$.
The multiplication by the window function corresponds to the convolution of the delta function with the FT of the window. It is simpler to consider a window function that is unity in the $(-T/2,T/2$ interval, as it is a real and even function, whose FT is also real and even (the $(0,T)$ case leads to an FT with the same amplitudes, but non-zero phases). In this case, it is simple to calculate that

\begin{equation}
    F(w(t)) = 2 {\sin(\pi\nu T)\over\pi\nu}
    \label{eq:sinc}
\end{equation}

which is a {\it sinc} function. One can see two windows of duration $T=1$s and $T=5$s and their FT in Fig. \ref{fig:window_ft}. The FT peak is broader for shorter $T$.

Therefore, the effect of a finite window, something that cannot be avoided in an astronomical observation, is the broadening of narrow peaks. The resolution of the signal FT is therefore higher the longer the observation is. In addition to the broadening, there is the formation of side lobes. They are much lower than the central peak, but cannot always be ignored. In the PDS, the drop in peak power scales as $\nu^{-2}$, so if the signal contains a noise component that is steeper than that, the power {\it spilled over} to higher frequencies due to the right-side side lobes is more than the signal power at those frequencies. As a result, the final slope will be -2, the slope of the envelope of the side lobes. In other words, no signal steeper than $\nu^{-2}$ in power can be recovered using a Fourier transform. 

\subsubsection{Sampling effects: aliasing}

As to the effect of sampling, the FT of a series of regularly spaced delta functions with spacing $T/N$ is itself a series of delta functions with spacing $T/N$, as shown in Eqn. \ref{eq:sampling}.

\begin{equation}
    s(t)    = \sum\limits_{k=-\infty}^{+\infty} \delta(t - {kT\over N}) \Longleftrightarrow
    F(s(t)) = \sum\limits_{m=-\infty}^{+\infty} \delta(\nu - {mN\over T})
    \label{eq:sampling}
\end{equation}

\begin{figure}
	\includegraphics[width=\columnwidth]{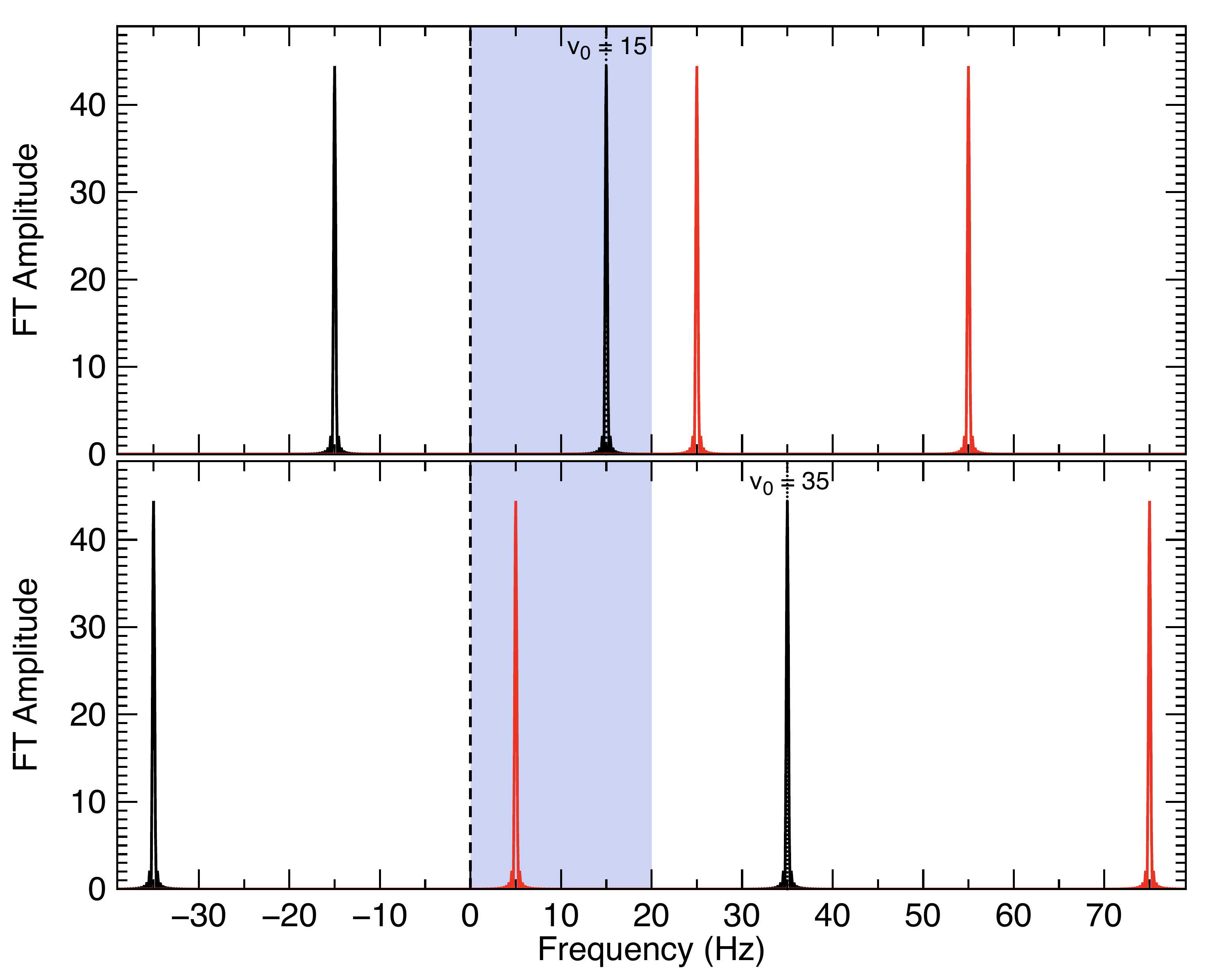}
    \caption{Top panel: Aliasing on a sinusoidal signal at 15 Hz sampled at 40 Hz. In black the true signal FT amplitude, in red the aliased one. The blue region marks the interval below $\nu_{Nyq}$. The signal is detected and the aliases are not.
    Bottom panel: same as the top panel, but with a signal at 35 Hz. Here the signal is not detected, but the 5 Hz alias is.
    }
    \label{fig:aliases}
\end{figure}

Therefore the effect of sampling on the FT of a sinusoidal signal with frequency $\nu_0$ (a delta function at $\nu_0$) is that of adding an infinite sequence of delta functions spaced by $N/T$, called {\it aliases}. What is important is that $N/T$ is twice the Nyquist frequency (see Eqn. \ref{eq:nyquist}). This ensures that if the data contain a sinusoidal signal at a frequency below $\nu_{Nyq}$, the aliases will not be in the ``allowed'' frequency range $(0,\nu_{Nyq})$. Notice that the FT amplitude is an even function for a real function, so the aliases will also be present in the negative frequency range. An example can be seen in the top panel of Fig. \ref{fig:aliases}). Here we have a signal at $\nu_0=15 Hz$, which appears as two peaks, at $\nu_0$ and $-\nu_0$ (in black). The Nyquist frequency is $\nu_{Nyq}=20 Hz$, therefore the region where we can search with our FT is that marked in blue. Because of aliasing, the two peaks are also repeated infinitely in both directions, with a step equal to $2\nu_{Nyq} = 40 Hz$, two of which can be seen in the plot (red). These aliases do not represent a problem, as they are both beyond $\nu_{Nyq}$.

However, problems arise when the signal is at a frequency {\it above} $\nu_{Nyq}$, as in the bottom panel of Fig. \ref{fig:aliases}), where $\nu_0=35 Hz$. We are not in the condition of detecting this signal, but because of data sampling one of the aliases (in red) appears at $\nu_{a} = 5 Hz$, which means we see it in our analysis.

\begin{figure}
	\includegraphics[width=\columnwidth]{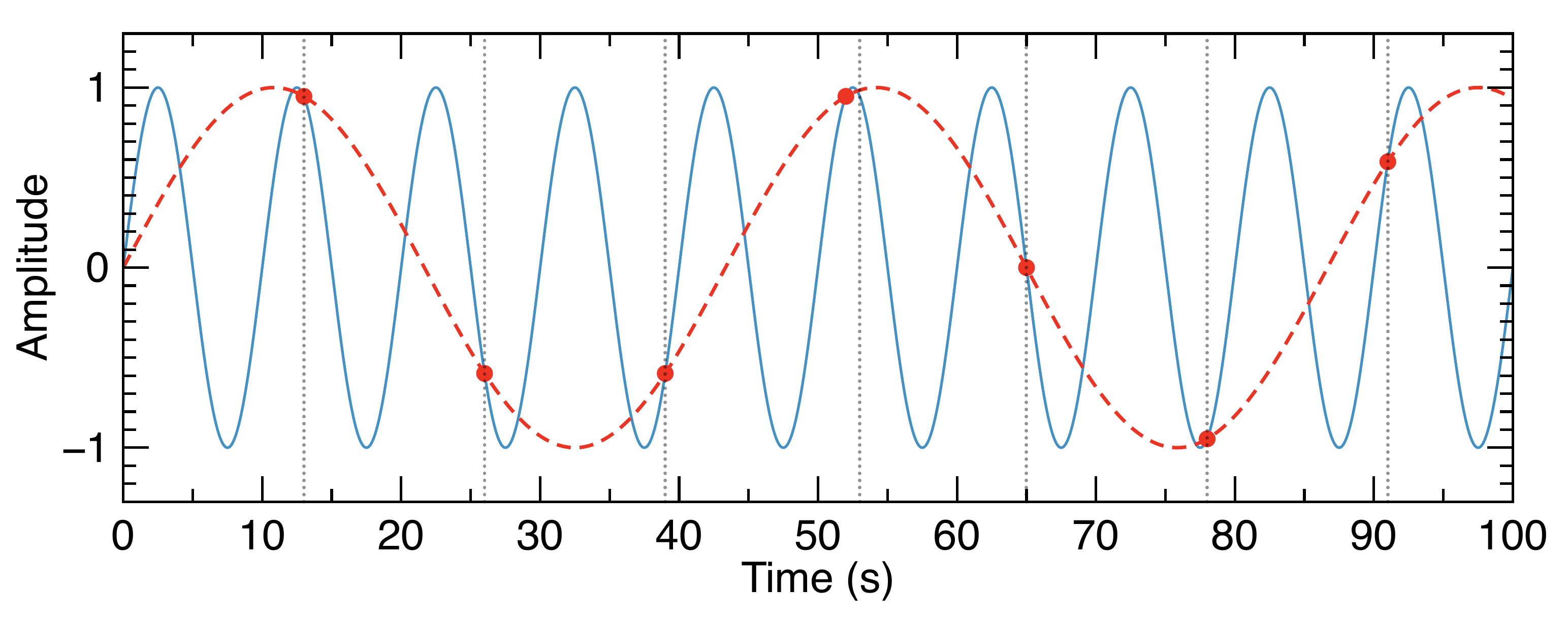}
    \caption{Time domain example of aliasing. The blue signal has $\nu_0 = 0.1$ Hz. If it is sampled with $\nu_{Nyq} = 0.038$ Hz (red points) the red dashed alias is the best fit to the data, with $\nu_a = 0.023$ Hz.
    }
    \label{fig:sampling}
\end{figure}

We have all experienced aliasing effects when looking at fast rotating objects like an air fan under fluorescent light. The light provides a sampling at 50 Hz (or 60 Hz, depending on where you live), while the fan has a periodicity. Depending on its angular speed, you can see it rotating apparently much slower, or even to stop and rotate in the opposite direction.
A graphical example in the time domain can be see in Fig. \ref{fig:sampling}. Here the signal (in blue) has a period of 10 s ($\nu_0 = 0.1$ Hz), but the sampling (red points) is every 13 s ($\nu_{Nyq} = 0.038$ Hz). The signal is out of range, but its alias at $\nu_a = 0.023$ Hz (period of 43.33 s, red dashed curve) is not.

This appears to be a very serious problem and it is, but in high-energy astronomy we do not sample signals, but integrate them over finite time bins. In this case, one does not multiply the signal by a sampling function $s(t)$, but {\it convolve} it with a binning function (see Eqn. \ref{eq:binning}).

\begin{equation}
b(t) = 
\begin{cases}
{N\over T} & t\in [-{T\over 2N},{T\over 2N}]\\
0                    & {\rm outside}
\end{cases}
\label{eq:binning}
\end{equation}

Therefore, the signal FT will be {\it mutiplied} by that of the binning function, which is again a {\it sinc} function and can be seen in Eqn. \ref{eq:binning2}.

\begin{equation}
   B(\nu) = {\sin \pi\nu /2\nu_{Nyq}\over \pi\nu /2\nu_{Nyq}}
\label{eq:binning2}
\end{equation}

$B(\nu)$ is a broad function that reaches 0 at $2\nu_{Nyq}$ and has the value of $2/\pi$ at $\nu_{Nyq}$. Therefore, aliasing is not an issue here. 
However, because of the reduction of amplitude due to the binning function, it is important to use, if possible, a fast binning, so that the Nyquist frequency is much higher than the signal one is looking for.

\subsubsection{Window carpentry}

We have seen the effects of the observing window upon the output FT and PDS. Of course such a window cannot be avoided, as one cannot have measurements infinite in duration. Having a longer observation reduces the width of the main window peak, but does not change the possible spillover effects. 
However, in some cases it can be advantageous to multiply the data by another window, not boxcar-shaped. This results in a loss of signal, as some data are multiplied by a factor less than unity, but there are advantages, depending on the chosen window.

\begin{figure}
	\includegraphics[width=\columnwidth]{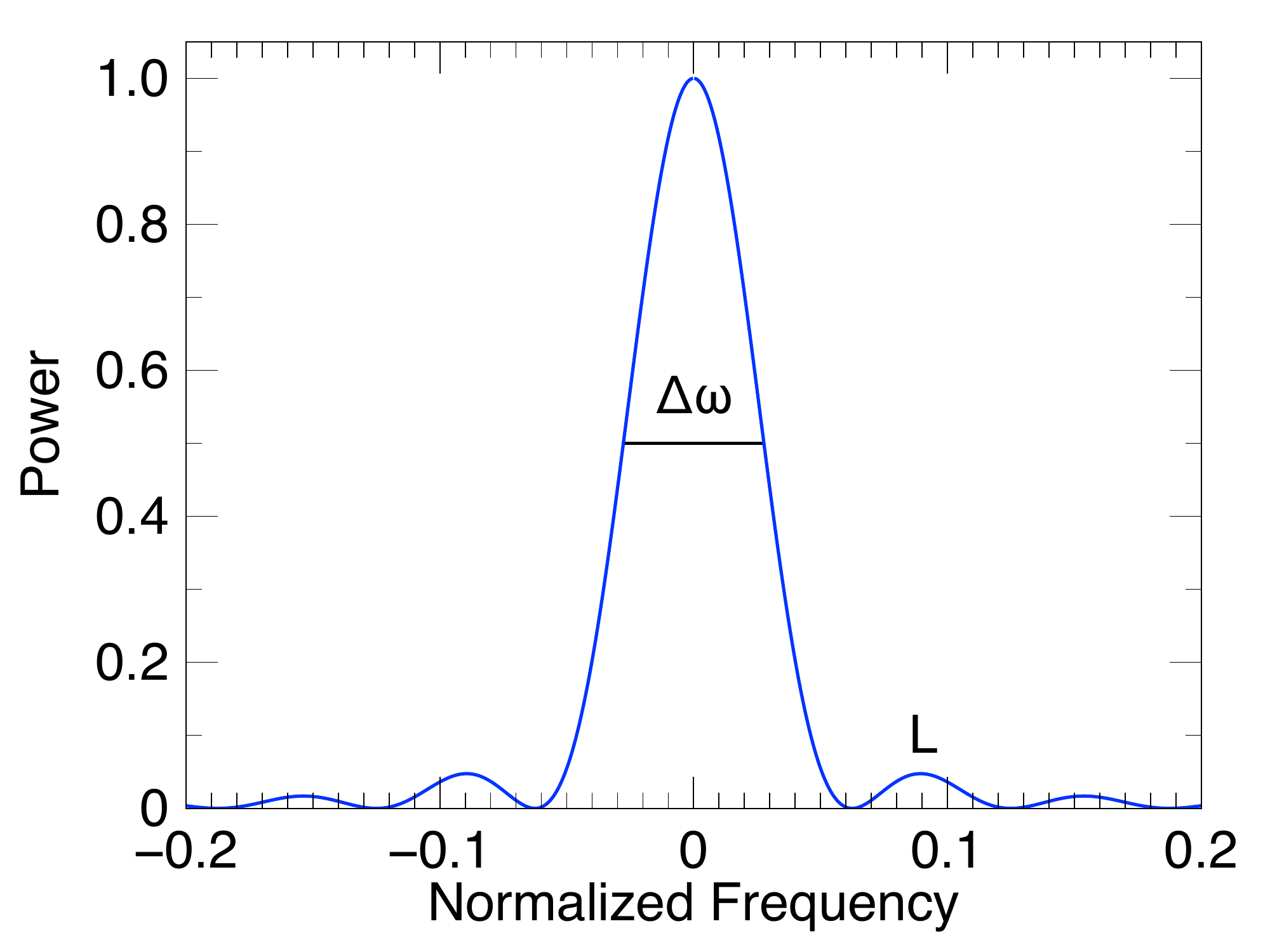}
    \caption{General shape of the PDS of a window function. Two main parameters are marked: $\Delta\omega$ and $L$ (see text). 
    }
    \label{fig:window_features}
\end{figure}

The above has led to the concept of {\it window carpentry}: many window functions with different characteristics have been designed and one can tailor them depending on what is needed. The main features that identify a window in its PDS are: the width of the main peak $\Delta\omega$, the relative amplitude of the first side lobe $L$ (expressed in decibels) and the slope of the decay of side lobes $n$ (see Fig. \ref{fig:window_features}).

\begin{figure}
	\includegraphics[width=\columnwidth]{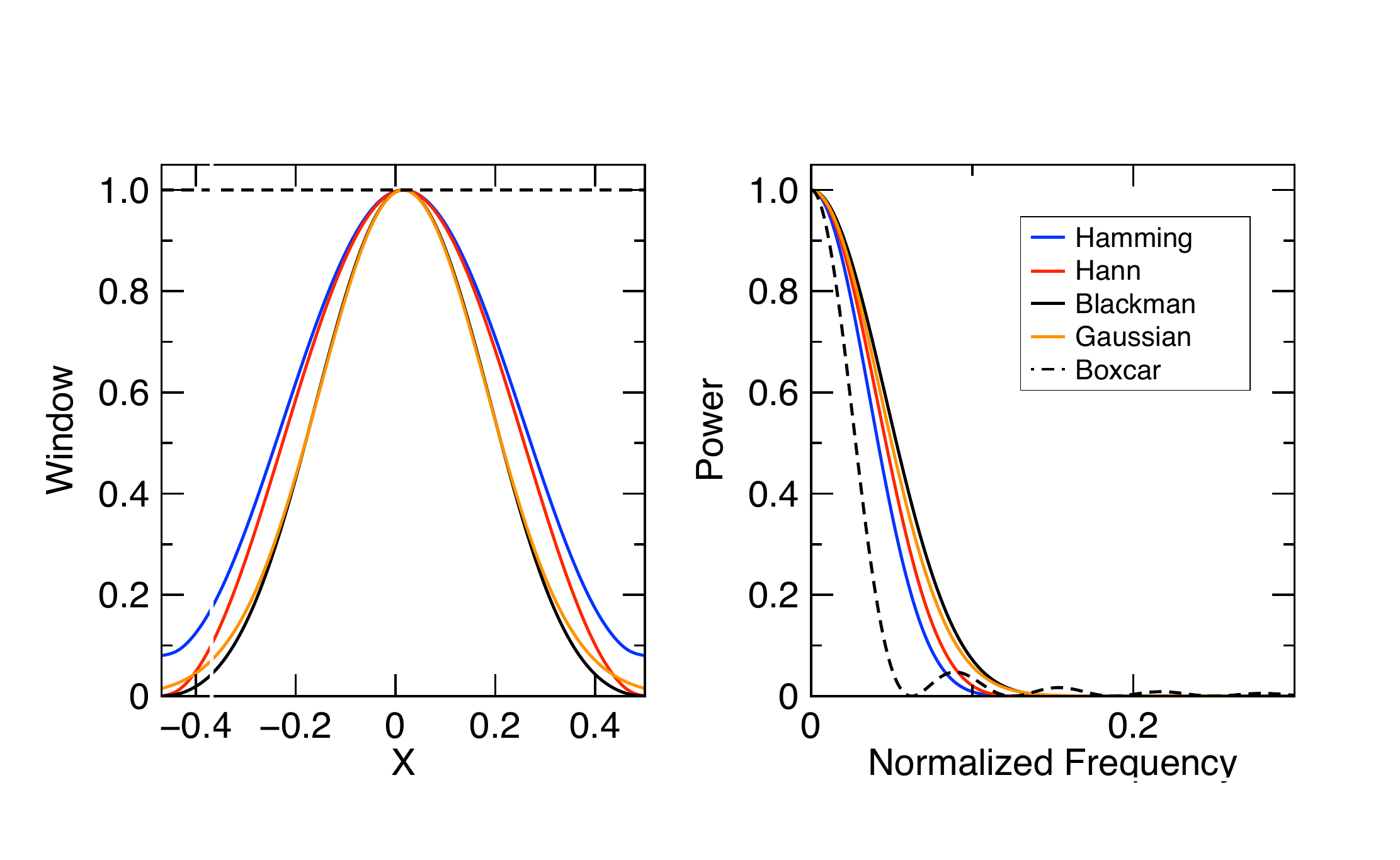}
    \caption{Left panel: the shape of five windows: boxcar, Hamming, Hann, Blackman and Gaussian. Right panel: the corresponding PDS, corresponding to (half) the shape of the PDS of a sinusoidal signal. The sidelobes of all but the boxcar window are too low to be seen (see Tab. \ref{tab:windows}).
    }
    \label{fig:windows}
\end{figure}

The boxcar window, the one you do not need to apply explicitly, is the one with the lowest $\Delta\omega$, but with alternative windows it is possible to obtain a significant reduction of the amplitude of the side lobes. The parameters of a few windows can be seen in Tab. \ref{tab:windows}, while the shape of those windows and their PDS are shown in Fig. \ref{fig:windows}. Many more windows have been proposed (see e.g. \cite{butz}).

\begin{table}
	\centering
	\caption{The functional shape and main PDS parameters of five windows}
	\label{tab:windows}
	\begin{tabular}{|c|c|c|c|c|} 
		\hline
		Window    & $\Delta\omega$ & L & n & Function\\
		\hline
        Boxcar    &    0.89         & -13db & 2 & 1\\
        Hamming   &    1.36         & -43db & 2 & $0.54+0.46\cdot\cos{(2\pi t)}$\\
        Hann      &    1.44         & -32db & 5 & $0.5\cdot[1-cos(2\pi t)]$\\
        Blackman  &    1.68         & -58db & 5 & $0.42+0.5\cdot\cos(2\pi t) + 0.08 \cdot\cos(4\pi t)$\\
        Gaussian  &    1.55         & -56db & 2 & $exp(-4.2*x^2)$\\
		\hline
	\end{tabular}
\end{table}

\subsubsection{Observational windows}

\begin{figure}
	\includegraphics[width=\columnwidth]{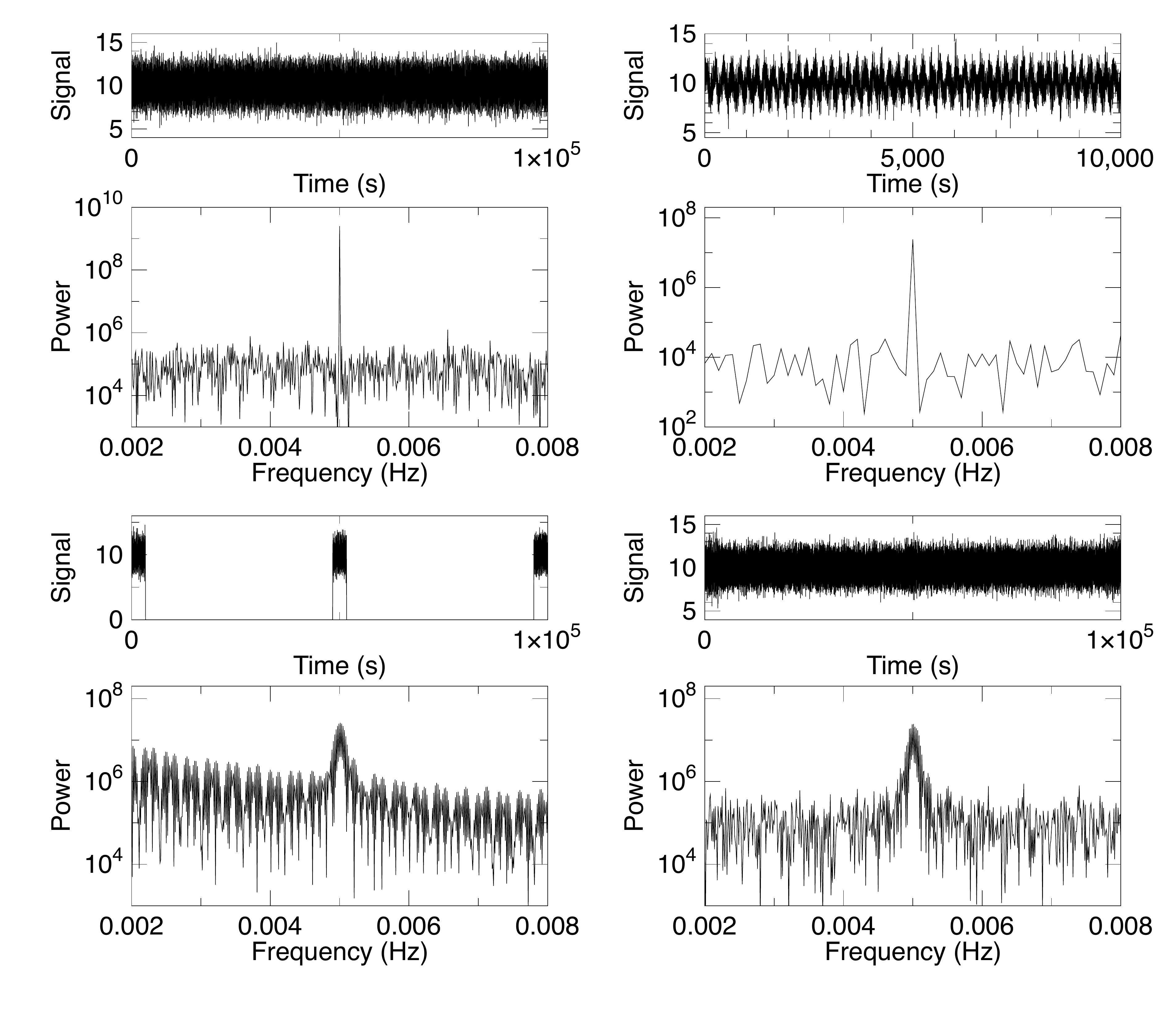}
    \caption{Each of the four panels contains a signal (top) and its PDS (bottom). In all four, the signal consist of Gaussian noise plus a sinusoid at P=200 s, with 1-s binning.
    Top left: continuous exposure of $10^5$ s.
    Top right: continuous exposure of $10^4$ s.
    Bottom left: exposure of $10^4$ s split into three intervals. The PDS is computed including the gaps as consisting of points at zero level.
    Bottom right: exposure of $10^4$ s split into three intervals as in the previous case. The gaps are filled with Gaussian noise.
    }
    \label{fig:window_comparison}
\end{figure}

As we have seen, even if no custom window is applied to the data, a boxcar window is determined by the start and end time of the signal. In case the signal is made of separate intervals, the production of a single PDS over the full time span means that a multiple boxcar window is applied. This results in a more complicated effect on the output PDS.
An example can be seen in Fig. \ref{fig:window_comparison}. 
In the top left pair of panels is a time series of $10^5$ s, with 1-s binning, containing Gaussian noise and a sinusoid with a period of 200 s and its PDS, zoomed to the frequencies around 0.005 Hz. The oscillation appears as a narrow peak, due to the long duration of the signal. 
In the top right panel the signal is the same, but limited to $10^4$ s: the peak in the PDS is broadened.
In the bottom left panel the signal is the same, but split into three 3333 s intervals distributed at equal distances over $10^5$ s. The PDS was made including the gaps as zero points. The effect of the more complex window is visible.
In the bottom right panel the two large gaps are filled with Gaussian noise with the same average as the signal. The window effects on the PDS are reduced, as the sharp drops are removed, but the broadening of the peak, together with its sidelobes, remains.

\subsection{Fast Fourier Transform}

Evaluation of the Discrete Fourier Transform (DFT, eqn.~\ref{eq:Timing:Basic:DFT}) of $N$ samples involves $\sim N^2$ multiplication and addition operations -- for every Fourier component $a_j$, each of the $N$ samples $x_k$ needs to be multiplied by a phase factor $e^{-2\pi ijk/N}$ and then they have to be summed.  The Fast Fourier Transform algorithm has been devised to accomplish this computation in much fewer steps, typically with $\sim N\log_2 N$ multiplications and additions.  This provides enormous computational savings for large transforms, and has made Fourier analysis accessible to cases where it would have been otherwise prohibitive.  Several different versions of the FFT algorithm exist.  We will illustrate the concept using the original algorithm of Cooley and Tukey \cite{cooleytukey} for transforms of length $N$ equalling integer powers of two, as presented in \cite{press_etal}. One may write the Discrete Fourier Transform (DFT) as
\begin{equation}
    a_j=\sum\limits_{k=0}^{N-1} W^{jk}x_k, \;\;\;\;\;\; W\equiv e^{-2\pi i/N}
\end{equation}
This can then be divided into even and odd parts:
\begin{eqnarray}
    a_j & = & \sum\limits_{k=0}^{N/2-1} e^{-2\pi ij(2k)/N}x_{2k}+\sum\limits_{k=0}^{N/2-1} e^{-2\pi ij(2k+1)/N}x_{2k+1} \\
    & = & \sum\limits_{k=0}^{N/2-1} e^{-2\pi ijk/(N/2)}x_{2k}+W^j\sum\limits_{k=0}^{N/2-1} e^{-2\pi ijk/(N/2)}x_{2k+1} \\
    & = & a_j^e+W^ja_j^o
\end{eqnarray}
This is a sum of two transforms, each requiring $\sim (N/2)^2$ operations, and thus the total number of operations required is reduced by a factor of two.  One may now continue to divide each of these transforms further, gaining a factor of two reduction in computation at each step.  Taking this all the way down to one-point transforms (identity operations), the net number of required operations to construct the full transform becomes $\sim N\log_2 N$.  In the implementation of the algorithm, some additional bookkeeping is required regarding which element of the original array needs to be combined with which others at each stage of the transform.  It turns out that if in the beginning the elements of the original array are reordered in a special way, then each step of the computation can be carried out by operations involving  just the adjacent elements or the adjacent transform products.  To do such a rearrangement, for any element of the original array one first expresses the array index in binary notation.  One then reverses the order of the bits of the index value to generate a new index, to which the element is now moved.

\section{The Power Density Spectrum and its representation}

As we have seen, the PDS is a powerful way of representing a signal in the frequency domain, both when one is interested in coherent signals and when the data contain incoherent noise originating from the source. In this section, we discuss the choice for PDS normalization, which is important both for statistical reasons and for extracting physical information. Moreover, the way the PDS is represented is also important in order to highlight important information.

\subsection{PDS Normalization}
Since the FT is a linear transformation and the PDS is its squared modulus, the PDS scales with the square of the intensity level of the signal (see Parseval's theorem above).  It is possible to normalize the PDS in different ways. In high-energy astronomy, two normalizations are commonly used, each of which has a different purpose.
The first normalization was introduced by \cite{Leahy}:

\begin{equation}
    P_j^{Leahy} = {2\over N_\gamma} |a_j|^2
    \label{eqn:leahy}
\end{equation}

where $N_\gamma$ is the total number of photons in the signal. This normalization leads to a known statistical distribution of signal power: if the signal is dominated by fluctuations due to Poisson statistics and if $N_\gamma$ is large, powers follow a chi square distribution with 2 degrees of freedom, $<P> = 2$ and $Var(P) = 4$. 
The reason is that the periodogram is the sum of the squares of the real and imaginary parts of the FT. For a stochastic process, the latter are normally distributed, so the sum of their squares is distributed as a chi square with 2 degrees of freedom.
If the signal is divided into $S$ segments and the resulting PDS are averaged (see below) and rebinned by a factor $M$, the powers will be distributed as a chi square with $2SM$ scaled by $1/SM$: therefore, the average power remains $<P> = 2$, but the variance is now $Var(P) = 4/SM$ (see also \cite{vdk88}). This is very important in order to establish the significance of an excess power over the (flat) Poissonian level. 

The so-called {\it Leahy normalization} however does not allow a direct extraction of quantitative indicators such as the fractional rms and does not remove the dependence of power on the signal intensity. A different normalization was introduced in 1990 by \cite{bellonihasinger}, usually called {\it ``Belloni normalization''} or {\it ``rms normalization.''}, obtained dividing the power by the net source intensity:

\begin{equation}
    P_j^{Belloni} = {1\over(C-B)^2} |a_j|^2
    \label{eqn:belloni}
\end{equation}

where $C$ is the detected intensity (source plus background) and $B$ is the background intensity. The reason for this choice is not statistical, but to have a PDS in units of {\it squared fractional rms}. With this normalization, different observations or sources can be compared in terms of fractional rms. Computing the square root of the integral of the the power in a given frequency range yields directly the fractional rms in that range. In principle it would be possible to take the square root to convert it to fractional rms, but this would alter the shape of the PDS.

The Poissonian contribution to the PDS is therefore in principle expected to be a flat spectrum at the level of 2 and with variance 4. Since the noise due to counting statistics is, also in principle, independent of the intrinsic signal, the cross term in Eqn. \ref{eq:pds_sum} is null and the flat component can be subtracted from the PDS. After an estimate of the Poissonian component is obtained, it is usually subtracted from the PDS. However, detector dead time does modify the shape and the level of the Poissonian component and introduces a correlation between source and noise signals (this is discussed elsewhere in this book). 

\subsection{PDS representation}

\begin{figure}
	\includegraphics[width=\columnwidth]{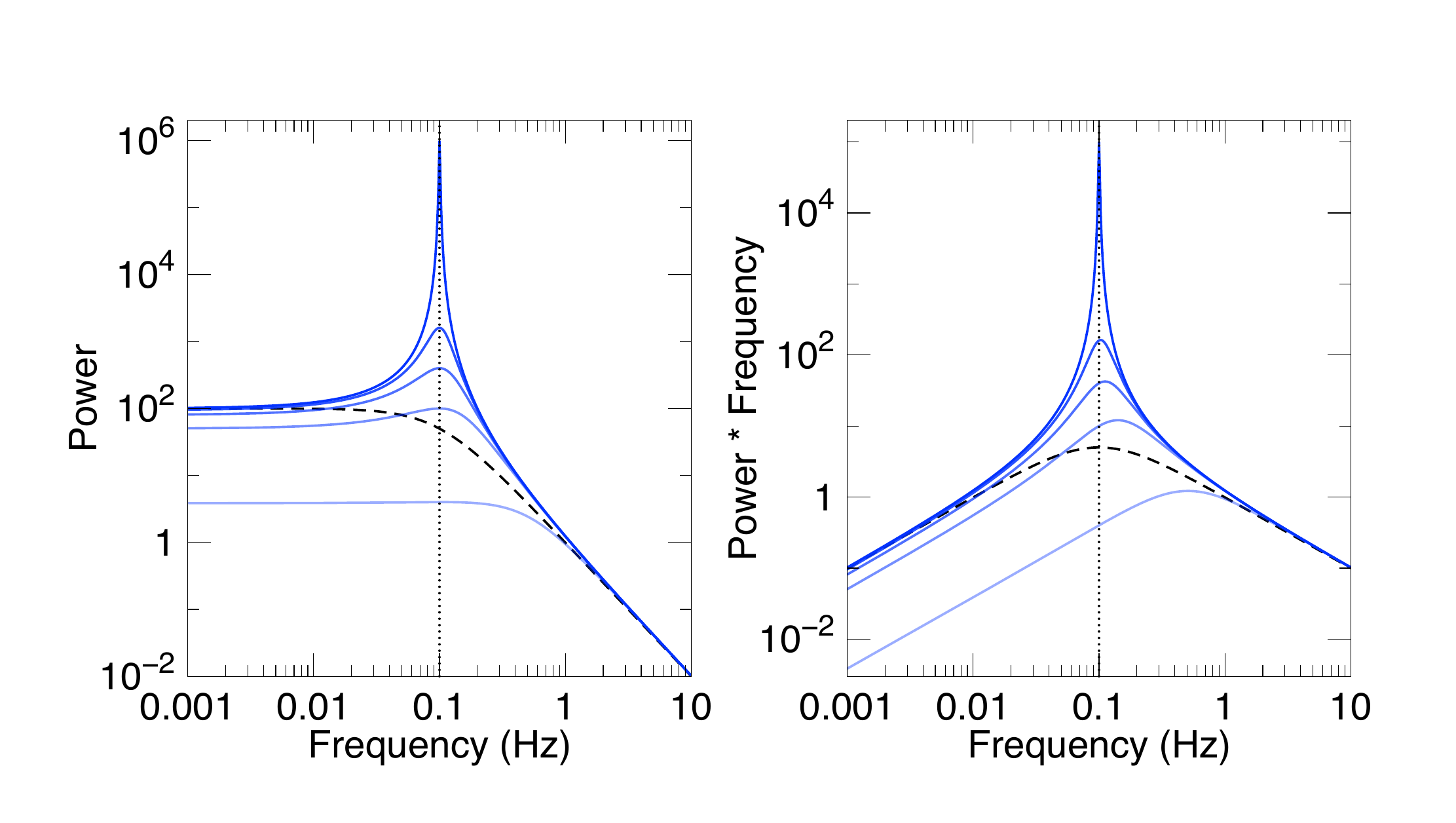}
    \caption{Left panel: power vs. frequency representation of PDS. The blue lines represent Lorentzian components with centroid $\nu_0=0.1 Hz$ (marked with a dotted line) and different Q values (from bottom to top: $Q=0.1,0.5,1.0,2.0,50.0$ (see text). The dashed black line show a zero-centered Lorentzian with $\Delta=0.1 Hz$. Adapted from \cite{bpk}.
    }
    \label{fig:representation}
\end{figure}

The PDS is usually plotted as a function of frequency, limiting the frequency axis to the allowed band between $1/T$ and $\nu_{Nyq}$. Examples are shown in the left panel of Fig. \ref{fig:representation}.
A different way to represent PDS graphically is now also being used (see \cite{belloni1997}, where the power on the y axis is multiplied by frequency, as it is done with energy spectra. Examples can be seen in the right panel of Fig. \ref{fig:representation}. This representation, called $\nu P_\nu$, shows squared rms per decade and therefore is more useful to assess the rms contribution of the different components. Moreover, in the $\nu P_\nu$ representation, Lorentzian functions peak at their $\nu_{max}$ frequency (see below). Finally, it is easier to visualise power-law behavior with indices between 0 and -2, as they become +1 and -1 respectively.

As we have seen, one way to reduce uncertainties in the PDS is rebinning in frequency. One has to be careful, as a coherent feature, especially in a long observation, can be very narrow and be diluted or even lost after rebinning. Moreover, for displaying broad noise components, it is common to rebin the PDS not linearly, but logarithmically. Instead of averaging $n$ bins all over the frequency range, each bin is made larger than the previous by a small amount; a typical value is around 1\%. This is advantageous because, unlike the case for coherent peaks, a broad component can emerge more clearly from the data after rebinning, since its power is not concentrated in a small number of bins.

\subsection{PDS decomposition}

As we have seen above, the PDS is not a linear transformation and is not additive. Cross terms can be neglected if there is no correlation between contributing components. Detector-related issues aside, Poissonian noise is uncorrelated from source signal and can be treated as an additive component. For separate source components, this is not necessarily true, in particular for source noise components, but it is customary to ignore the possibility of non-zero cross-terms.

\begin{figure}
	\includegraphics[width=\columnwidth]{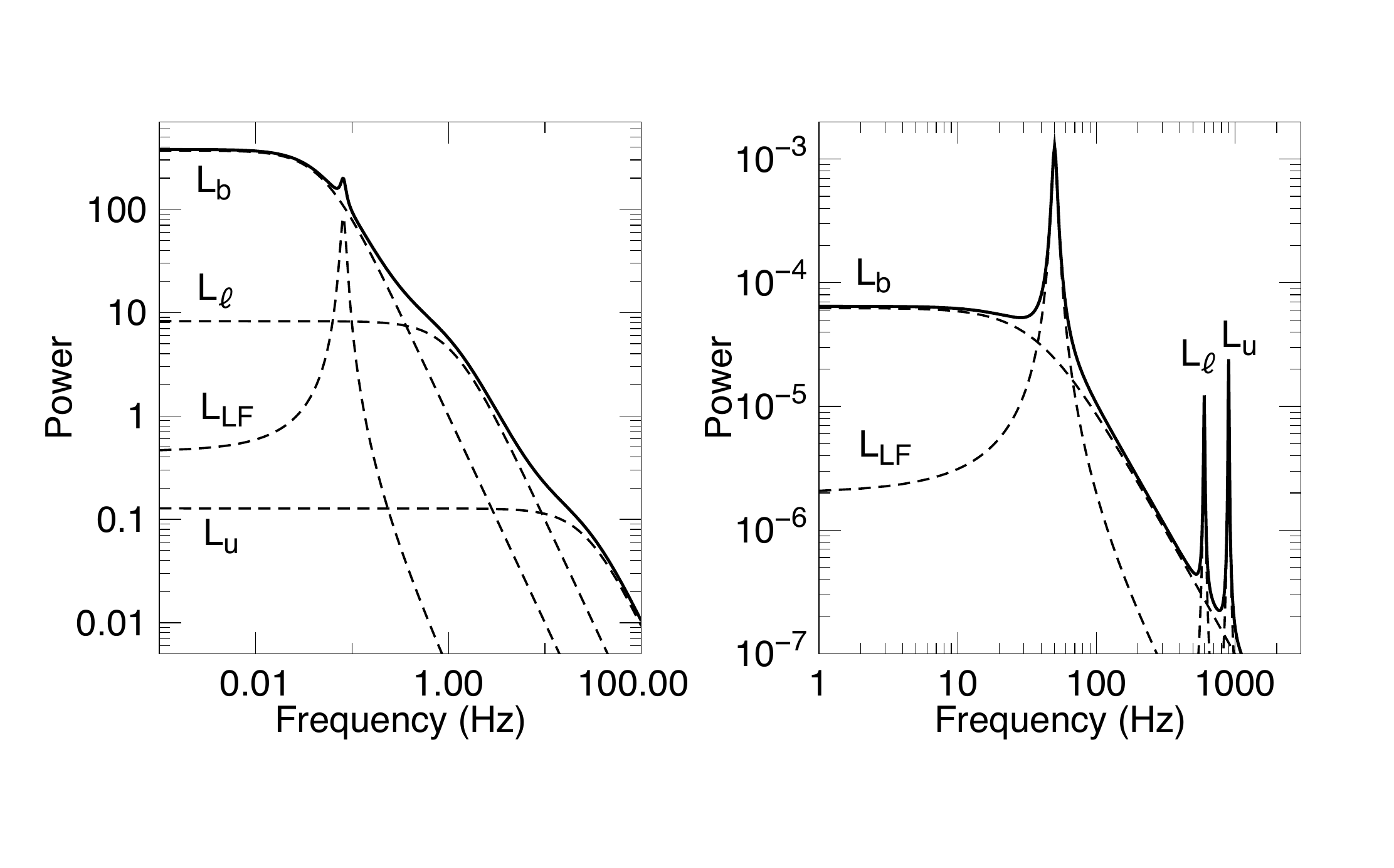}
    \caption{Left panel: Lorentzian decomposition of a typical PDS of a black-hole binary in hard state. Right panel: Lorentzian decomposition of a typical NS LMXB with kHZ QPOs. The four Lorentzian in each PDS have been proposed to be the same components, as indicated by the labels.
    Adapted from \cite{bpk}.
    }
    \label{fig:lorentzians}
\end{figure}

In the case of coherent oscillations such as pulsations from neutron stars, the shape of the PDS is determined by the window used (see above) and, in the case of binary systems, the smearing that results from orbital doppler effects. In many cases, the signal itself consists of complex and often strong noise components, which require a functional characterization. While in the past broken power laws have been used, it has become customary to fit these PDS with a combination of both broad and peaked components modeled as {\it Lorentzians} (see Eqn. \ref{eq:lorentzians}).

\begin{equation}
    P(\nu) = {r^2 \Delta\over\pi} {1\over \Delta^2 + (\nu-\nu_0)^2}
    \label{eq:lorentzians}
\end{equation}

where $\nu_0$ is the centroid frequency, $\Delta$ the HWHM and $r$ the integrated fractional rms. How coherent such a component is can be characterized by its {\it quality factor} $Q=\nu_0/2\Delta$ \cite{nowak,bpk}. In the case of $\nu_0 = 0$ we have a {\it zero-centered Lorentzian}, flat at low frequencies and decreasing as $\nu^{-2}$ at high frequencies, which is used to fit band-limited noise. Examples can be seen in Fig. \ref{fig:representation}.
The decomposition of noise PDS into a sum of homogeneous components has helped to unify timing properties of accreting X-ray binaries (see Fig. \ref{fig:lorentzians}), as Lorentzians can fit both narrow and broad components. In particular this decomposition has led to the identification of {\it characteristic frequencies}. For a Lorentzian, the characteristic frequency $\nu_{max}$ is defined as in Eqn. \ref{eq:characteristic} \cite{belloni1997,bpk}.

\begin{equation}
    \nu_{max} = \sqrt{\nu_0^2 + \Delta^2} = \nu_0 \sqrt{1+{1\over 4 Q^2}}
    \label{eq:characteristic}
\end{equation}

As can be seen in Fig. \ref{fig:representation}, $\nu_{max}$ is the frequency at which the Lorentzian contributes most in terms of power per logarithmic frequency and also the frequency at which the component peaks in the $\nu P_\nu$ representation. From Fig. \ref{fig:representation} and Eqn. \ref{eq:characteristic} one can see that, for a narrow component such as that with $Q=50$, $\nu_{max}\approx \nu_0$, but for broad components it deviates substantially and is more representative of a ``special'' frequency in the PDS.

However, although the fits are often good, it has to be noted that we do not yet have a physical backing behind the use of Lorentzian components, other than the generic fact that a Lorentzian is the PDS of a damped oscillator.

\subsection{Bartlett's method and data gaps}

As we have seen, with the {\it Leahy} normalization the noise power is distributed as a chi square with two degrees of freedom: this means that the average is 2 and the standard deviation is 2. Quite a noisy spectrum! We also have seen that dividing into $S$ segments and rebinning by a factor of $M$ reduces the error. The technique of dividing the time series into equal-duration intervals and averaging the corresponding PDS is called {\it Bartlett's method} and is commonly used in high-energy astronomy. Of course, this does not allow to detect changes in the variability properties with time for a non-stationary signal, which we will examine later. Also, the reduction in time duration $T$ increases the minimum frequency in the PDS $\nu_{min} = 1/T$.

This method also allows one to skip over data gaps, which have dramatic effects on the PDS. A gap in the time series corresponds to an interval of signal at level 0, which means a modification to the boxcar window, with serious effects on the resulting PDS. In addition to skipping the gaps, selecting only continuous intervals, one can (if the gaps are short) fill them with a simulated signal. For instance, an average value extrapolated from non-gap parts, adding Poissonian noise. However, the effects on the PDS will not be completely removed and gap-filling can add biases to the data.

\section{Auto and cross-correlation}

The Cross-correlation of two functions $f(t)$ and $g(t)$ is defined as
\begin{equation}
    C(\tau)=f\star g=\int\limits_{-\infty}^{\infty}f^*(t)g(t+\tau)dt
    \label{eq:Timing:Basic:Correlation:def}
\end{equation}
The result is a function of the ``lag'' $\tau$ introduced between the two functions, and is often used to estimate the similarity between two different time series, as a function of lag.  For example, if a common underlying process causes the time variation of intensity at two different electromagnetic bands but the signals suffer differential delays while propagating to the observer, then the cross correlation function of the two time series will exhibit a peak at the corresponding lag, namely the relative delay between the two bands.

The autocorrelation function (Eqn.~\ref{eq:acf}) is a special case where a function is correlated with itself,
which would always show a peak at zero lag.  Convolution (Eqn.~\ref{eq:convolution}) is an operation akin to the Cross-correlation, but the function $g(t)$ in the integrand is inverted to $g(-t)$ before adding the shift.

Unlike convolution, Cross-correlation is not commutative:
\begin{equation}
    [f\star g](\tau) = [g^*\star f^*](-\tau)
\end{equation}
Other important properties of Cross-correlation include:
\begin{eqnarray}
[f\star g]\star[f\star g] & = & [f\star f]\star[g\star g] \\
g\star(f\otimes h) & = & [g\star f]\otimes h \\
F[f\star g] & = & F(f)\cdot F^*(g) \label{eq:Timing:Basic:FTCrossCor}
\end{eqnarray}
where $F$ represents the Fourier transform.

The definition in Eqn.~\ref{eq:Timing:Basic:Correlation:def} involves the functions $f$ and $g$ over the entire real line.  In practical use, both these would be time series of finite duration, and not necessarily of equal length.  The input to the correlation integral will therefore not be the original functions $f$ and $g$ defined over $(-\infty,\infty)$ but windowed copies of them: $fw_f$ and $gw_g$ where $w_f$ and $w_g$ are boxcar window functions of amplitude unity over the durations of the respective time series and zero elsewhere.  As can be seen, this would cause a lack of overlap between parts of the two functions when they are sufficiently shifted with respect to each other.  There are two ways this can be handled -- the non-overlapped portion could either be wrapped, resulting in a {\em cyclic} correlation, or could be ignored.  Cyclic correlation is appropriate for periodic functions.  

In other cases, ignoring the non-overlapped portions will decrease the lengths of the functions being multiplied, and will thus alter the net normalisation of the integral.  To account for this, one may divide the integral by the Cross-correlation of the two window functions $w_f$ and $w_g$ at the same lag:
\begin{equation}
    C(\tau)=\frac{[fw_f\star gw_g](\tau)}{[w_f\star w_g](\tau)}
\end{equation}

For discretely sampled functions, this is equivalent to dividing by the total number of overlapped points at the corresponding lag:
\begin{equation}
    C(j)  =  \frac{1}{N_{\rm eff}}\sum\limits_{i=i_{\rm min}}^{{\rm Min}(N,M-j)} f_i\, g_{i+j}
\end{equation}
with $N_{\rm eff} = [{\rm Min}(N,M-j)-i_{\rm min}]$, $i_{\rm min}$ being zero for $j\ge 0$ and $-j$ for $j<0$.
Here $f_i,\, i=0,1,...N$ and $g_i,\, i=0,1,...M$ are the two discrete functions sampled at equal intervals. 

Often it is also customary to normalise the cross correlation function such that its amplitude does not exceed unity.  This is achieved by dividing the cross correlation by the square root of the product of the two autocorrelation functions at zero lag:
\begin{equation}
    C_{\rm N}(\tau)=\frac{C(\tau)}{\sqrt{[f\star f](0)[g\star g](0)}}
\end{equation}
The result is referred to as the Normalised Cross Correlation Function.

The Fourier transform of the Cross-correlation function defines the Cross Spectrum (Eqn.~\ref{eq:Timing:Basic:FTCrossCor}).  Non-zero lags in the cross correlation function naturally manifest as corresponding phase gradients in the cross spectrum.  It is easy to see that the autocorrelation function $ACF(f)$ has for its Fourier transform $|F(f)|^2$, namely the Power Spectrum. 

\begin{figure}
    \centering
    \includegraphics[width=\columnwidth]{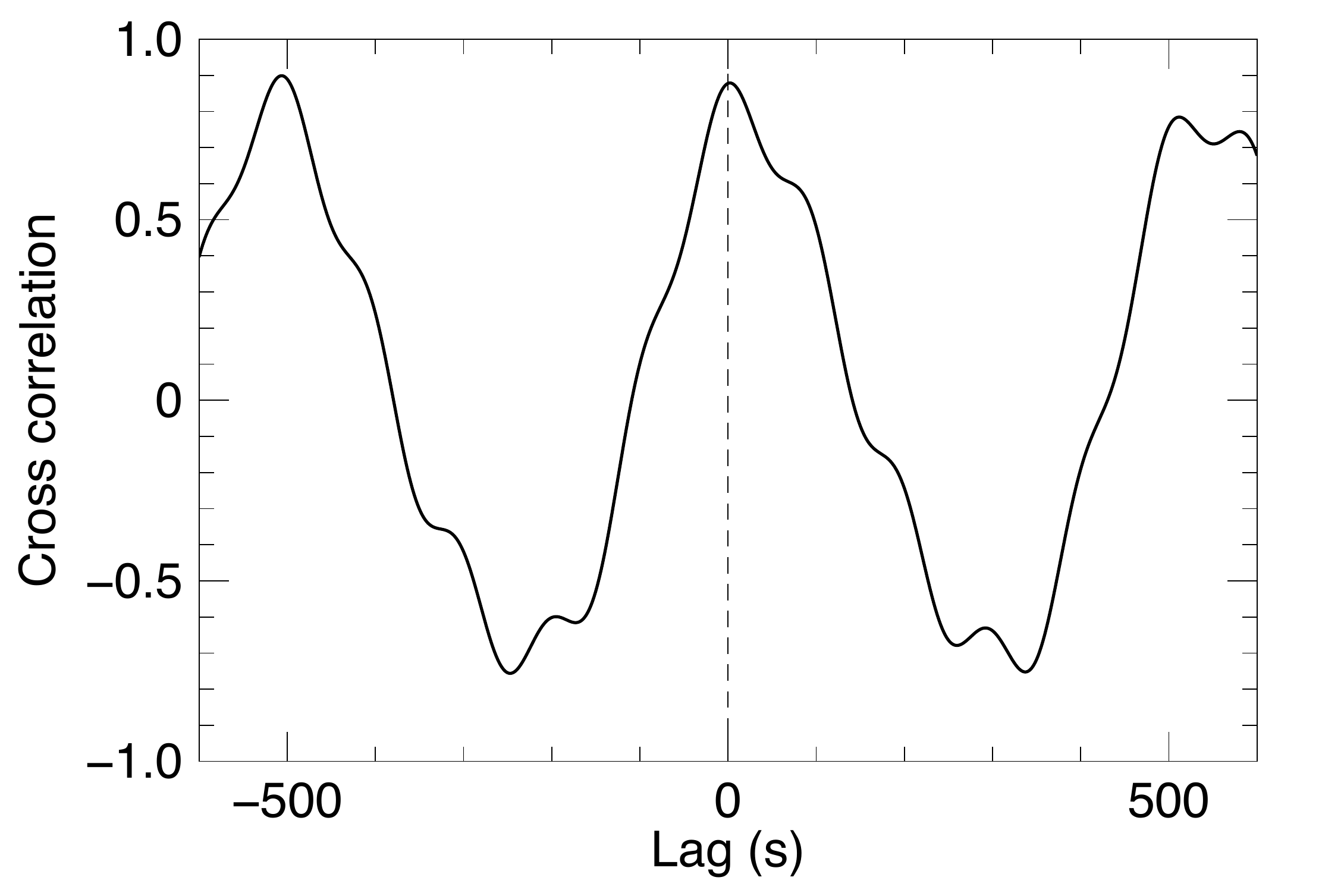}
    \caption{Normalised Cross Correlation Function of two simulated time series, both composed of three sinusoids of periods 10, 23 and 56 seconds with amplitudes of 10, 10 and 30 units respectively.  In the second time series the three components have phase offsets of $-0.2$, $-0.3$ and $+0.3$ radians respectively with respect to those in the first.  Despite these phase offsets, the Cross Correlation Function peaks close to zero lag. However, it has an asymmetric profile.}
    \label{fig:Timing:Basic:CrossCorrSim}
\end{figure}

At times, the observed time series could be composed of multiple narrowband components, with the lags for each of them being independent, and different.  Depending on the relative amplitudes of these components, such lags may not necessarily manifest themselves in the cross correlation function as distinct peaks offset from zero lag, but instead cause an asymmetry in the cross-correlation profile. Fig.~\ref{fig:Timing:Basic:CrossCorrSim} shows an example of this.  Such frequency-dependent phase lags are more clearly detected in the Cross Spectrum, as illustrated in the next section.

\section{Cross-spectra, phase lag spectra and coherence}

Above we have introduced the PDS as the Fourier Transform of a function times its complex conjugate, and we have shown that the PDS is the Fourier Transform of the autocorrelation function. Then, we have introduced the cross-correlation between two time series. As in the case of the autocorrelation, the cross-correlation does not allow to discriminate between different frequencies and is usually used only to assess global delays between simultaneous signals. 
Also in this case, it is possible to explore the frequency domain through the {\it cross spectrum}. Given two signals $f(t)$ and $g(t)$ and their respective FTs $F(\omega)$ and $G(\omega)$ we define the cross spectrum as in Eqn. \ref{eq:cont_cross_spectrum}:

\begin{equation}
CS(\omega) = F(\omega)\cdot G^*(\omega)
\label{eq:cont_cross_spectrum}
\end{equation}

\begin{figure}
    \centering
    \includegraphics[width=\columnwidth]{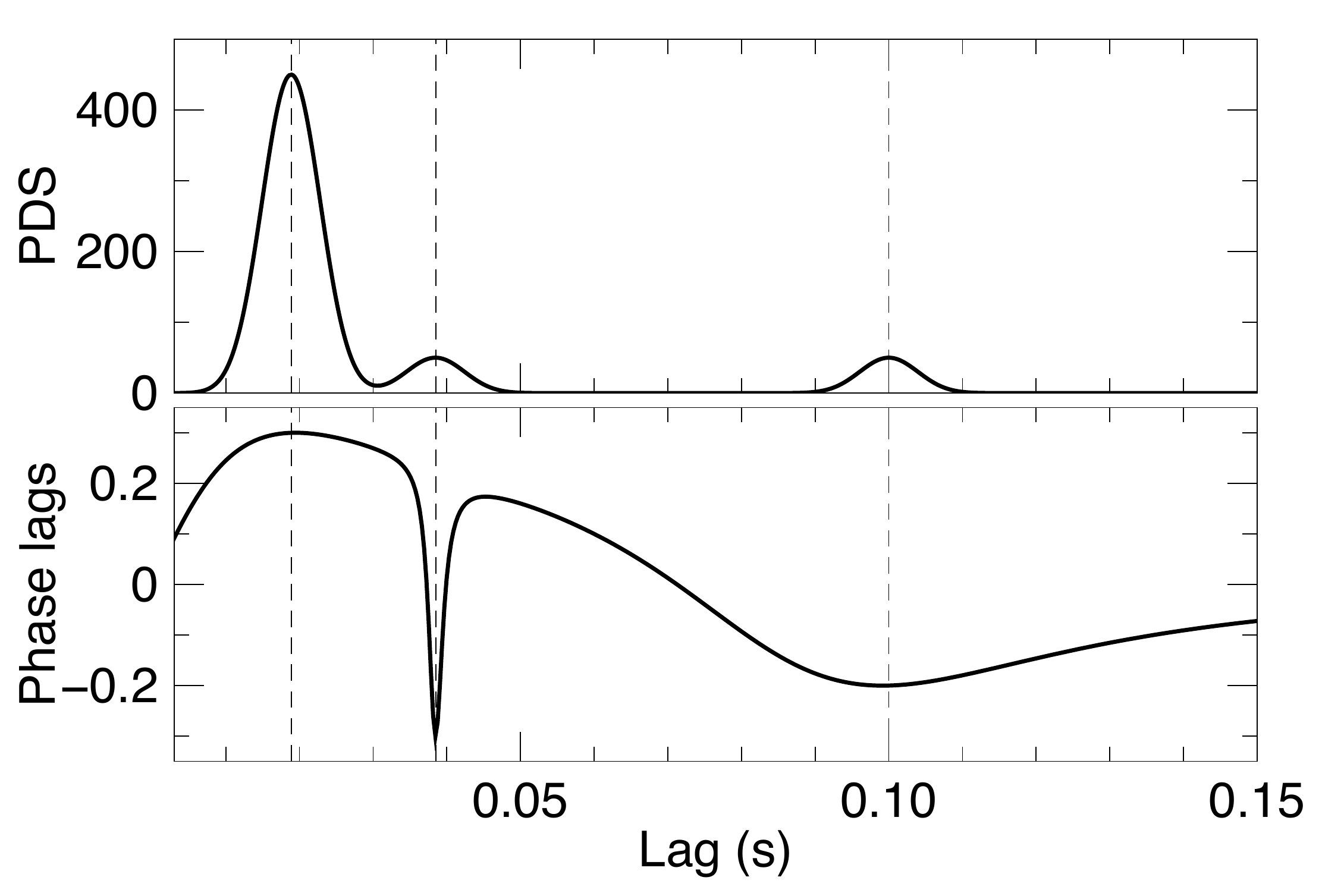}
    \caption{Top panel: PDS of the same simulated time series as in Fig.\ref{fig:Timing:Basic:CrossCorrSim}, consisting of three sinusoids of periods 10, 23 and 56 seconds. 
    Bottom panel: Phase lag Spectrum of the two simulated time series. The lags at the frequencies of the three sinusoids, marked as dashed lines, are very visible.}
    \label{fig:Timing:Basic:CrossSpectrumSim}
\end{figure}

Analogous to the PDS and the autocorrelation, the cross spectrum between two signals is the Fourier transform of their cross-correlation (and vice-versa). Since, unlike the autocorrelation, the cross-correlation is not by definition an even function, the cross spectrum is not by definition a real quantity. At each frequency, its argument represents the phase difference between intensity fluctuations in the two signals at that frequency. The difference in phase at a certain frequency can be translated into a time delay dividing it by the frequency.
As an example, Fig. \ref{fig:Timing:Basic:CrossSpectrumSim} shows the PDS of the signal for which the cross-correlation in Fig. \ref{fig:Timing:Basic:CrossCorrSim} was calculated, with the peaks corresponding to the three sinusoids, and the Cross Spectrum between the two time series with lagged sinusoids. In the cross spectrum, the lags of the three sinusoids are clearly readable. This Cross Spectrum is much easier to interpret than the corresponding cross-correlation.

While for a coherent signal the phase delays (or phase lags) are simple to interpret, in many cases the signal consists of noise plus broad peaks. For these signals, the interpretation is not directly obvious, as the decomposition of broad-band noise into sinusoids does not correspond to physical oscillations at the different frequencies. It is important to remark that the phase of a sinusoid at a certain frequency is a quantity that makes sense only for that frequency (or multiples of it), but it does not make sense to compare it with the phase at a different frequency. This means that averaging phases or phase lags over a range of frequencies does not result into a physical quantity. This does not apply of course to time lags.

As we have seen, the cross spectrum of two signals at each frequency is a complex number whose argument represents the phase delay between the signals at that frequency. 
However, let us suppose to have a set of independent measurements of the two signals, for instance obtained by slicing the signals into $n$ shorter samples. Then an important information is whether the phase delay measured with the cross spectrum is the same across samples. This can be measured by computing the {\it coherence} between the two signals. If the two signals are $s_1(t)$ and $s_2(t)$, their PDS are $P_1(\nu)$ and $P_2(\nu)$ respectively and their cross spectrum is $CS(\nu)$, then their coherence is defined as in Eqn. \ref{eq:coherence}

\begin{equation}
\gamma^2(\nu) = {|<CS(\nu)>|^2 \over
                <P_1(\nu)> <P_2(\nu)>}
\label{eq:coherence}
\end{equation}

where the angle brackets mean average over the $n$ samples.
The averaging of the cross-spectra at the numerator in Eqn. \ref{eq:coherence} means that at each frequency the complex vectors are summed. If their argument, representing the phase delay at that frequency, is always the same, the resulting sum is a straight vector, if their are random it will be a null vector. Intermediate cases will lead to a non-straight vector. The coherence represents the ratio of the actual sum vector and the straight vector. In case of constancy of phase delays, it will be unity. In case of random phase delays it will be zero. In other words, if the two signals are related by a linear transform, the coherence is unity at all frequencies.


\section{Bispectrum and bicoherence}

If $f(t)$ is a time series of measurements and $F(\omega)$ is its Fourier Transform, then we have seen above that the Power Spectrum is the Fourier Transform of the autocorrelation function of $f(t)$.  The autocorrelation function may be thought of as the second order correlation
\begin{equation}
    c_2(\tau)=\langle f(t)f(t+\tau) \rangle
\end{equation}
where the angular brackets denote an ensemble average.  Ideally the average so defined would involve infinite length of data train $f(t)$, but in practice even with shorter lengths the average would be independent of time $t$ if stationarity is assumed.

The Power Spectrum
\begin{equation}
    P(\omega)=|F(\omega)|^2\Longleftrightarrow c_2(\tau)
\end{equation} 
may then also be shown to be equal to
\begin{equation}
    P(\omega)=\langle F(\omega)F^*(\omega) \rangle=\langle F(\omega)F(-\omega) \rangle
    = \langle F(\omega)F(\omega^{\prime})\delta(\omega+\omega^{\prime}) \rangle
\end{equation}
the angular brackets again denoting ensemble average, this time in frequency domain.  The two frequencies $\omega$ and $\omega^{\prime}$ sum to zero as a consequence of stationarity.

Bispectrum is an extension of the above concept to triple correlations.  The third order correlation function
\begin{equation}
    c_3(\tau_1,\tau_2)=\langle f(t)f(t+\tau_1)f(t+\tau_2)\rangle
\end{equation}
would have the double Fourier Transform
\begin{equation}
    B(\omega_1,\omega_2)=\int\limits_{-\infty}^{\infty}\int\limits_{-\infty}^{\infty} c_3(\tau_1,\tau_2) e^{-i(\omega_1\tau_1+\omega_2\tau_2)}\,d\tau_1\,d\tau_2
\end{equation}
which is defined as the Bispectrum.  It can be shown that
\begin{eqnarray}
    B(\omega_1,\omega_2)& = & \langle F(\omega_1)F(\omega_2)F(\omega_3)\delta(\omega_1+\omega_2+\omega_3)\rangle \nonumber\\
    & = &\langle F(\omega_1)F(\omega_2)F^*(\omega_1+\omega_2)\rangle
\end{eqnarray}
The bispectrum provides information about non-linear interaction between waves. The bispectrum is non-zero at a pair of frequencies ($\omega_1$, $\omega_2$) only if the Fourier component $F(\omega_1+\omega_2)$  is statistically dependent on the product $F(\omega_1)F(\omega_2)$, indicating a non-linear coupling between the original frequencies or a specific phase relation between them.  

The bispectrum may be normalised to define a quantity called {\em bicoherence} $b(\omega_1, \omega_2)$, with a value between 0 and 1:
\begin{equation}
    b^2(\omega_1, \omega_2)=\frac{|B(\omega_1,\omega_2)|^2}{\langle |F(\omega_1)F(\omega_2)|^2\rangle\langle |F(\omega_1+\omega_2)|^2\rangle}
\end{equation}

\section{Lomb-Scargle technique for non-uniform sampling}

The Fourier methods described so far are designed to deal with evenly sampled data.  However, if the available data samples have non-uniform intervals, a technique that is often used in Astronomy to search for periodicities is the Lomb-Scargle periodogram.  An useful exposition of this method is presented in \cite{vanderplas2018}.  

As discussed above, the sampling function may be represented as a series of delta functions located at the sample points which, in this case, are non-uniformly spaced. The available time series is then a multiplication of the underlying function with this sampling function.  In the Fourier domain, the transform of this time series will be a convolution of the transform of the underlying function with that of the sampling function.  The latter, unlike in the case of uniform sampling, is noise-like in character.  The resulting convolution thus makes it harder to identify the features of the underlying function, hampering the search for periodicity.  Nevertheless, for sufficiently strong periodic signals, one can find associated peaks in the analogue of the Power Spectrum:
\begin{equation}
P(\omega)=\frac{1}{N}\left|\sum_{n=1}^{N}f_ne^{-i\omega t_n}\right|^2
\end{equation}
where $f_n$ are the time series values sampled at the epochs $t_n$, and $N$ is the total number of samples.  The above can be written as
\begin{equation}
P(\omega)=\frac{1}{N}\left[\left(\sum_{n=1}^{N}f_n\cos(\omega t_n)\right)^2+\left(\sum_{n=1}^{N}f_n\sin(\omega t_n)\right)^2\right]
\end{equation}
This is called the Classical or the Schuster
Periodogram.  The Lomb-Scargle Periodogram is a slight modification of this \cite{scargle82}:
\begin{eqnarray}
    P_{\rm LS}(\omega)& = &\frac{1}{2}\left\{ \left(\sum_nf_n\cos(\omega[t_n-\tau])\right)^2/\sum_n\cos^2(\omega[t_n-\tau])\right. \nonumber \\
    & & +\left.\left(\sum_nf_n\sin(\omega[t_n-\tau])\right)^2/\sum_n\sin^2(\omega[t_n-\tau]) \right\}
\end{eqnarray}
where 
\begin{equation}
    \tau=\frac{1}{2\omega}\tan^{-1}\left(\frac{\sum_n\sin(2\omega t_n)}{\sum_n\cos(2\omega t_n)}\right)
\end{equation}
This form is identical to that obtained by least-square fitting a model consisting of simple sinusoids at each frequency and constructing a periodogram out of the respective $\chi^2$ values \cite{lomb76}.  This modification of the expression of the classical periodogram simplifies some of its statistical properties so that relatively simple expressions can be used to estimate detection thresholds and false alarm rates \cite{scargle82}.

We have discussed above (Eqn.~\ref{eq:nyquist}) the role of sampling in limiting the highest frequency that the data are sensitive to, and referred to it as the Nyquist limit.  One property of an unevenly sampled time series is that the effective Nyquist limit may often be raised to high values, even well beyond the reciprocal of the shortest sampling interval present in the data.  In this context, Eyer and Bartholdi \cite{eyerbartholdi} prove the following: 
\begin{quote}
{\em Let $p$ be the largest value such that each sampling epoch $t_i$ could be written as $t_i=t_0+n_i p$, where $n_i$ are integers. Then the Nyquist frequency is $1/(2p)$.}  
\end{quote}
This reduces to the conventional Nyquist frequency in the case of uniform sampling.  In the case of non-uniform sampling one needs to find the largest interval $p$ such that each sampling interval present in the data is an exact integral multiple of $p$.  If the sampling intervals are truly incommensurate with each other then $p$ would be vanishingly small and there would be no Nyquist limit.

\begin{figure}
    \centering
    \includegraphics[width=\columnwidth]{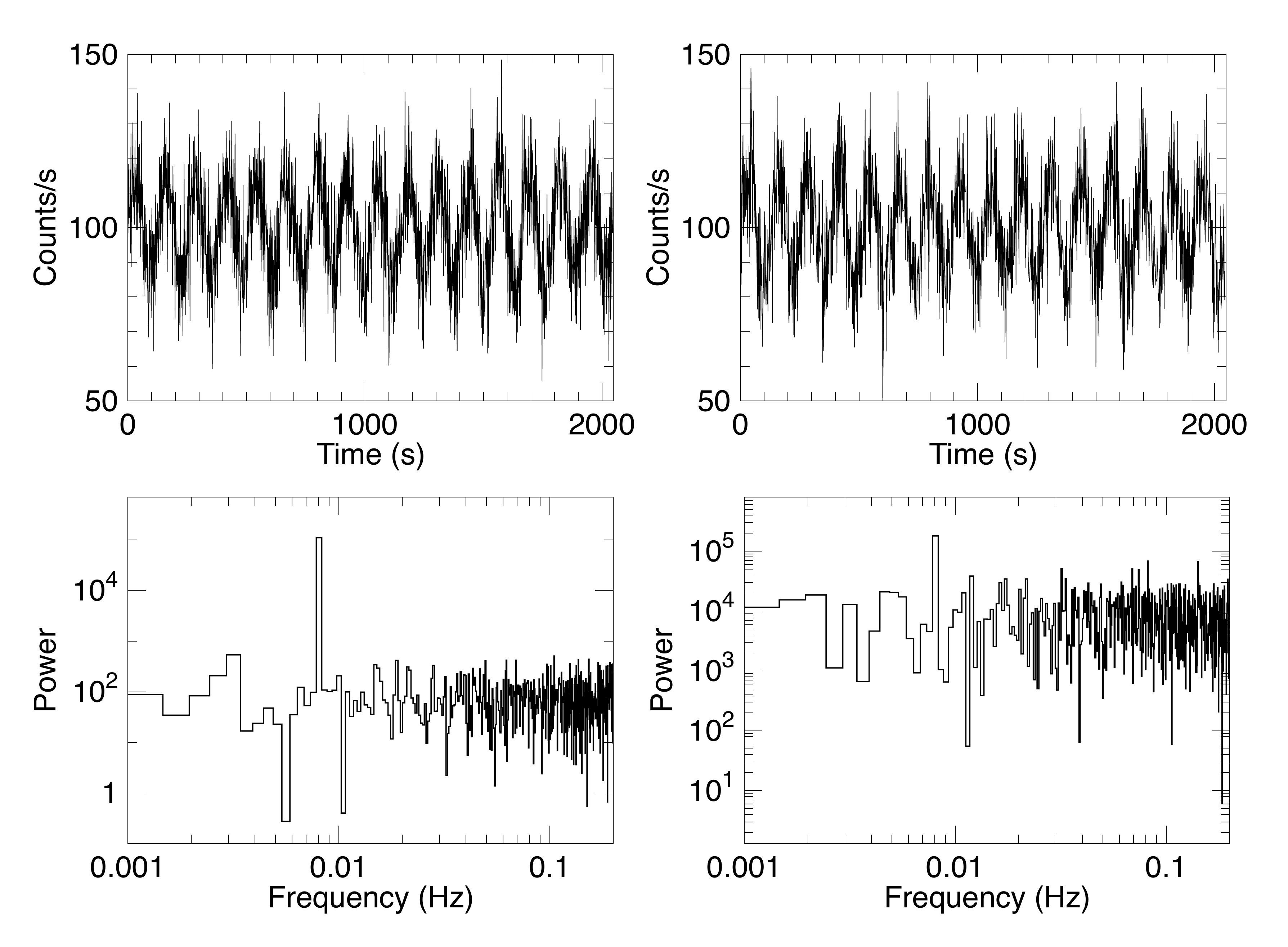}
    \caption{A simulated light curve (top) and its Lomb-Scargle periodogram (below).  On the left, the light curve is sampled uniformly at 1 second interval and on the right the sampling interval is random.  Both light curves are of the same duration (2048~s) and have the same total number of samples (2048).  The light curve consists of a sinusoid of period 128~s and amplitude 15 count~s$^{-1}$, riding on a gaussian background noise of mean 100~s$^{-1}$ and rms 10~s$^{-1}$. The periodicity is detectable in both periodograms as a spectral peak at 0.0078~Hz, but on the right the noise level is enhanced due to non-uniform sampling. The periodogram on the left is identical to the Power Spectrum obtained by Fourier Transform.}
    \label{fig:Timing:Basic:LombScargle}
\end{figure}
\section{Time-frequency analysis}

The properties of the observed signal are often not constant in time, i.e. the signal might not be stationary. For instance, there could be a state transition causing an abrupt change in the signal, or the characteristic frequency of a component could vary in time. Of course one can produce PDS from selected intervals of data in order to isolate different behaviours, but it is often more efficient to perform {\it time-frequency analysis}. 

\subsection{Short-time Fourier transform}

\begin{figure}
	\includegraphics[width=\columnwidth]{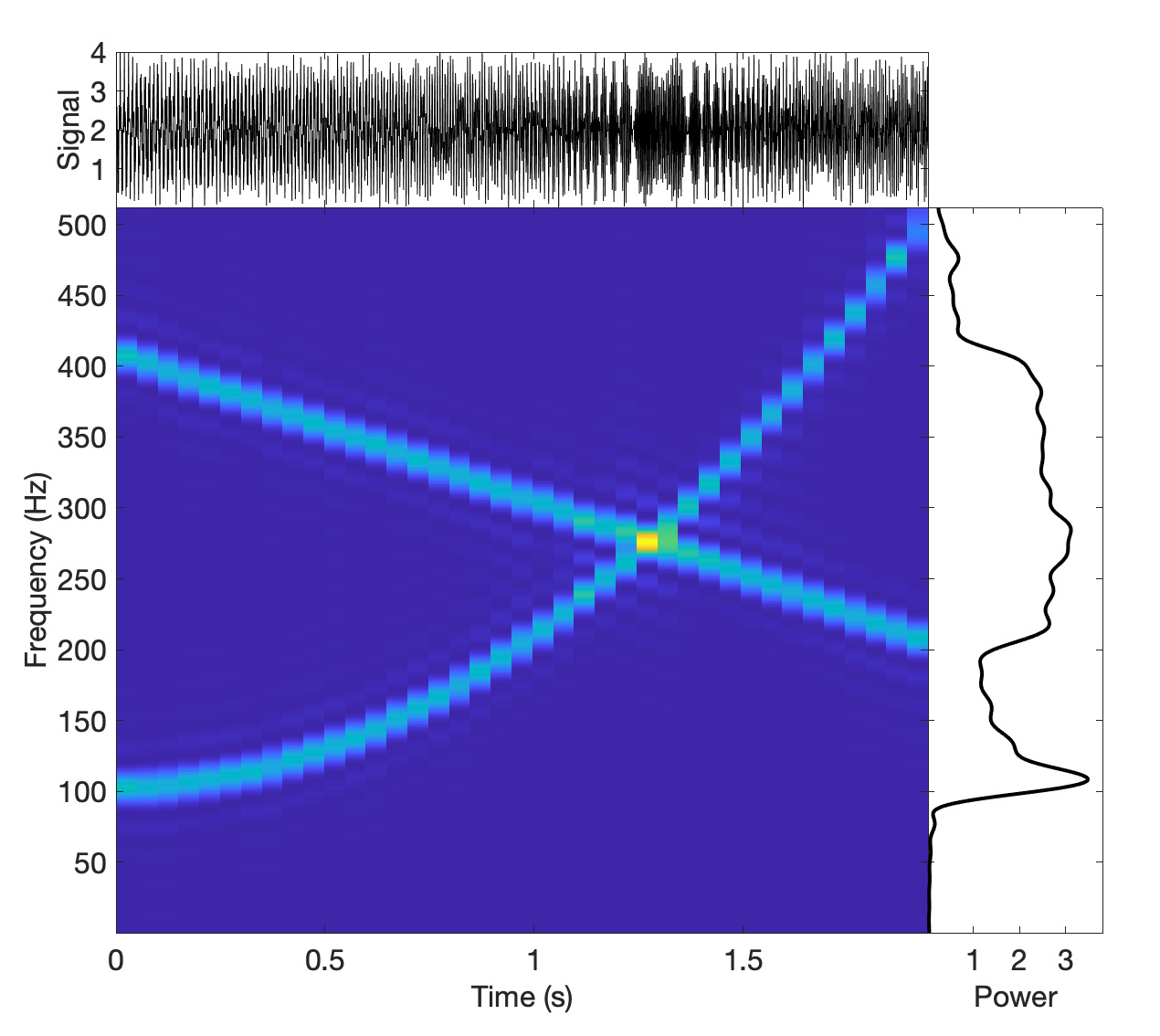}
    \caption{Top panel: time series containing two chirps. Right panel: corresponding Bartlett PDS. Central panel: spectrogram where the two chirps are clearly visible.
    }
    \label{fig:spectrogram}
\end{figure}

We have seen Bartlett's method, which consists in dividing the signal into equal-length intervals and averaging the corresponding PDS. Instead of averaging, we can keep the single FFTs (what is called {\it short-time Fourier transform}), generate PDS and produce a {\it spectrogram} (often called ``dynamical power spectrum" in high-energy astronomy): an image that contains power as a function of frequency and time. A simulated example can be seen in Fig. \ref{fig:spectrogram}: the signal is shown on top, the Bartlett PDS on the right and the spectrogram in the center panel. The signal is made of two ``chirps'', one whose frequency grows with time and the other having the opposite trend: this is not visible in the time series and PDS plots, but appears evident in the spectrogram. Notice the presence of side lobes, caused by the boxcar window, they can be suppressed using another window, at the expense of broadening the two chirp lines. 

We have shown above that the frequency resolution of the PDS of a signal of duration $T$ is $\Delta\nu = 1/T$ and this is also the resolution in frequency of the spectrogram. The resolution in time is the window duration $\Delta T = T$. This immediately shows that there is a trade-off between time and frequency resolution. This is an expression of the {\it Gabor limit}, which is itself connected to the {\it uncertainty principle} in time series analysis (indirectly related to Heisenberg's). There is no need to go into precise definitions and more mathematical detail, but this is a strong limitation for time-frequency analysis: the resolution element in the spectrogram cannot be made arbitrarily small in one direction without increasing it in the other. 

It is possible to divide the time series into overlapping intervals, so that each of them shares a percentage of time with the next one. This reduces frequency resolution, but also averages out noise. Adding the PDS results in the {\it Welch's method}, an extension of Bartlett's.

\subsection{Wavelets}

In the past decades, new mathematical tools have been developed for the analysis of both images and time series, called {\it wavelets}. For time series, they constitute a way to work around the Gabor limit. 

A wavelet is a ``small wave,'' where small derives from the fact that it is mostly limited to an interval of time. A wavelet $\psi$ has to satisfy two requirements: its integral must be zero and the integral of its square must be unity (see Eqn. \ref{eq:wavelet}).

\begin{equation}
\int\limits_{-\infty}^{\infty} \psi(t) dt = 0\quad\quad\quad\quad \int\limits_{-\infty}^{\infty} \psi^2(t) dt = 1
\label{eq:wavelet}
\end{equation}

\begin{figure}
	\includegraphics[width=\columnwidth]{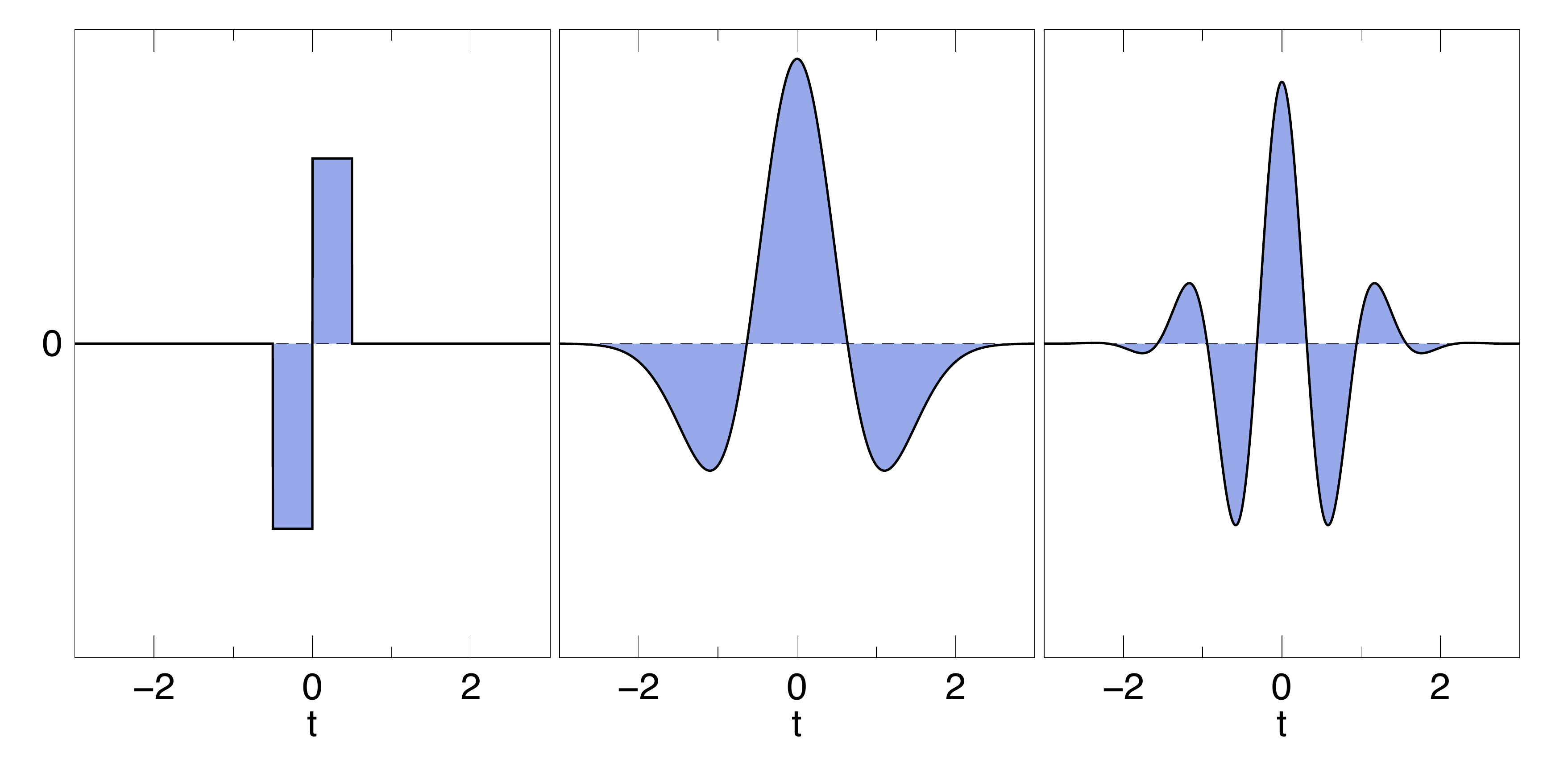}
    \caption{Three examples of wavelets. Left panel: Haar Wavelet. Middle panel: mexican hat wavelet. Right panel: Morlet wavelet.
    }
    \label{fig:wavelets}
\end{figure}

It is easy to see that these requirements imply that $\psi (t)$ is essentially non-zero only over a limited range of $t$ and that it has to extend both above and below zero. Three examples of wavelets can be seen in Fig. \ref{fig:wavelets}.

A wavelet can be shifted by $\tau$ and dilated by a scale parameter $\sigma$ (Eqn. \ref{eq:scaling}):

\begin{equation}
\psi_{\tau,\sigma}(t) = {1\over\sqrt{\sigma}}\psi\left({t-\tau\over\sigma}\right)
\label{eq:scaling}
\end{equation}

There are two types of wavelet transforms: continuous and discrete. The continuous transform, in which the scale and shift parameters are changed smoothly and can assume all values, is what can be used for time-frequency analysis. The discrete wavelet transform, where scales and shifts are changed in discrete fashion by factors of two, is used for compression and de-noising and has no use in the analysis of time series. Here, we will only deal with the continuous transform.
The wavelet transform of a function $f(t)$ is computed by correlating $f(t)$ with the complex conjugate of $\psi_{\tau,\sigma}(t)$ (Eqn. \ref{eq:wav_trans}) (wavelets can also be complex functions):

\begin{equation}
W[f(\tau,\sigma)] = \int\limits_{-\infty}^{\infty}  f(t) \psi^*_{\tau,\sigma}(t) dt
\label{eq:wav_trans}
\end{equation}

\begin{figure}
    \centering
    \subfloat[\centering Spectrogram]{{\includegraphics[width=\columnwidth]{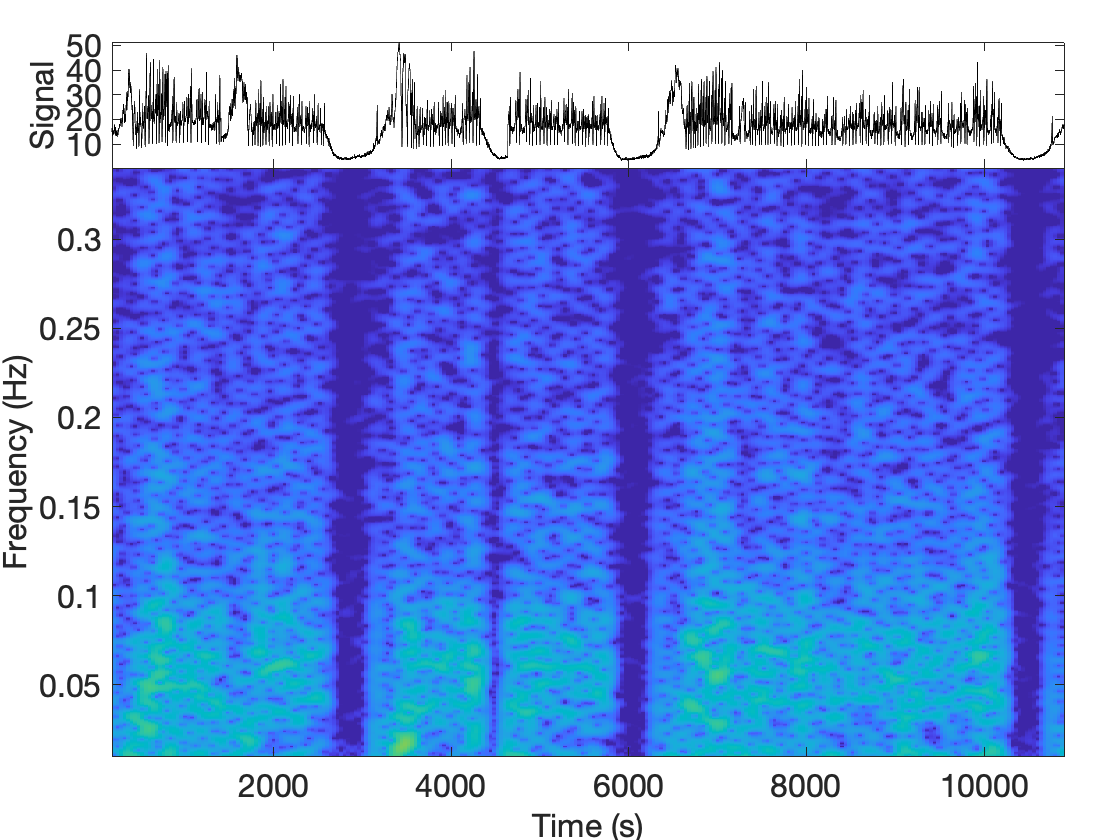} }}\\
    \subfloat[\centering Wavelet transform]{{\includegraphics[width=\columnwidth]{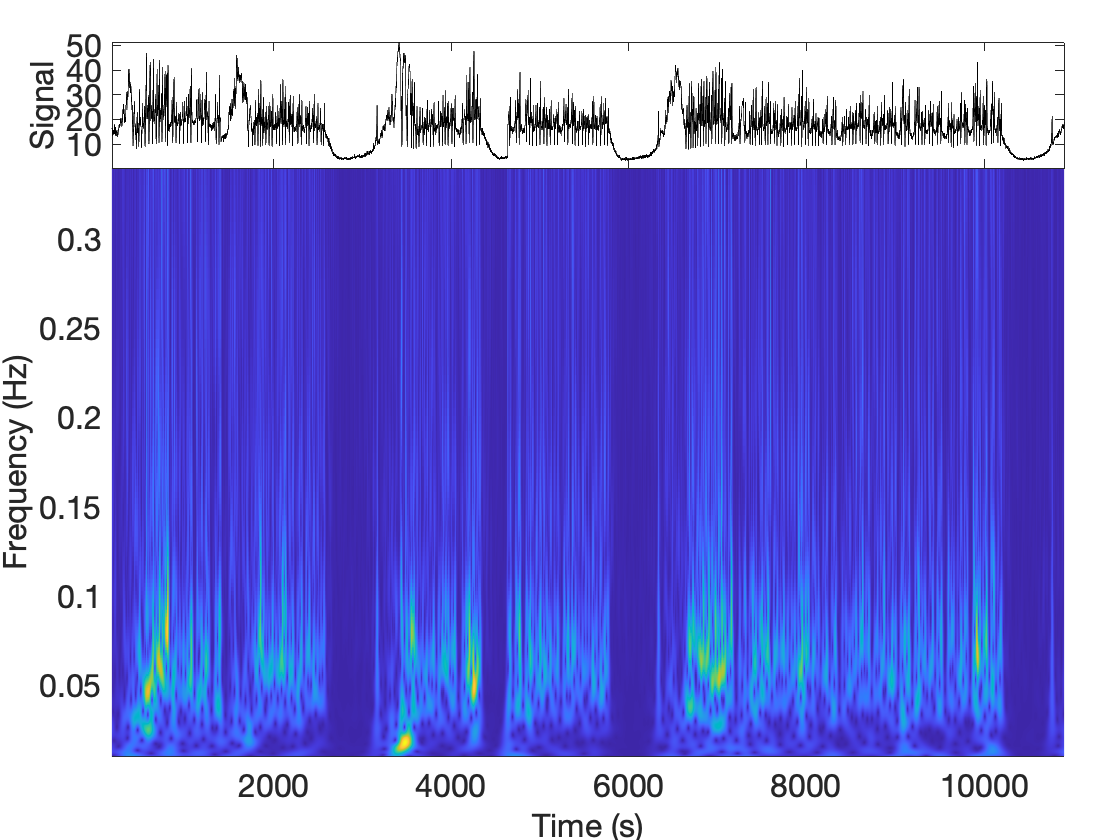} }}%
    
    \caption{(a) Top panel: an RXTE/PCA light curve of GRS 1915+105, with bin size 1s and gaps removed (in kcts/s). Bottom panel: corresponding spectrogram with time intervals of 400 s, 90\% overlapping and a Hann window. Power is in log scale.
    (b) Top panel: the same light curve as in (a). Bottom panel: corresponding amplitude (in log scale) of the wavelet transform using an analytic Morlet wavelet (a complex version of the Morlet wavelet). Frequencies are converted from scale factors for comparison.
    }
    \label{fig:grs1915}
\end{figure}

The power of the wavelet transform is that it allows localization both in time and frequency. Its properties allow to sample large time durations for low frequencies, and at the same time short time durations for higher frequencies by the scaling properties of the wavelet transform. An example is shown in Fig. \ref{fig:grs1915}. Here a very variable 1-s light curve of the bright and peculiar black-hole binary GRS 1915+105 is shown. The signal consists of intervals showing a strong quasi-periodicity alternated with broad smooth dips.
In panel (a) the spectrogram is shown (here with a Hann window and 400-s sliding intervals that overlap by 90\% with the previous one). Here one can see power in the 50-100 mHz range and absence of power in the dips, but not much can be seen in detail.
In panel (b) the wavelet transform is shown. A complex-valued version of the Morlet wavelet was used and what is shown is the amplitude of the complex valued transform (phases are also available). Wavelet scales have been converted to frequency for comparison (frequencies are inversely proportional to scales). Features corresponding to quasi-periodicities are much more visible. A {\it scalogram} can also be produced, where the square modulus of the amplitude of the wavelet transform is used.

More information on wavelets and wavelet transforms can be found in \cite{addison}, \cite{percivalwalden} and \cite{mallat}.

\subsection{Other techniques}

Finally, it is interesting to show, briefly, another approach to time-frequency analysis. It starts from the {\it Wigner-Ville distribution} (WV), originally developed by Wigner for calculations in physics and brought to the signal analysis field by Ville. Given the usual time series $f(t)$, its Wigner-Ville distribution (in its continuous representation) is shown in Eqn. \ref{eq:wigner}):

\begin{equation}
WV(t,\omega) = {1\over 2\pi}\int f^*(t-{1\over 2}\tau) f(t+{1\over 2}\tau) e^{-i\tau\omega} d\tau
\label{eq:wigner}
\end{equation}

\begin{figure}
	\includegraphics[width=\columnwidth]{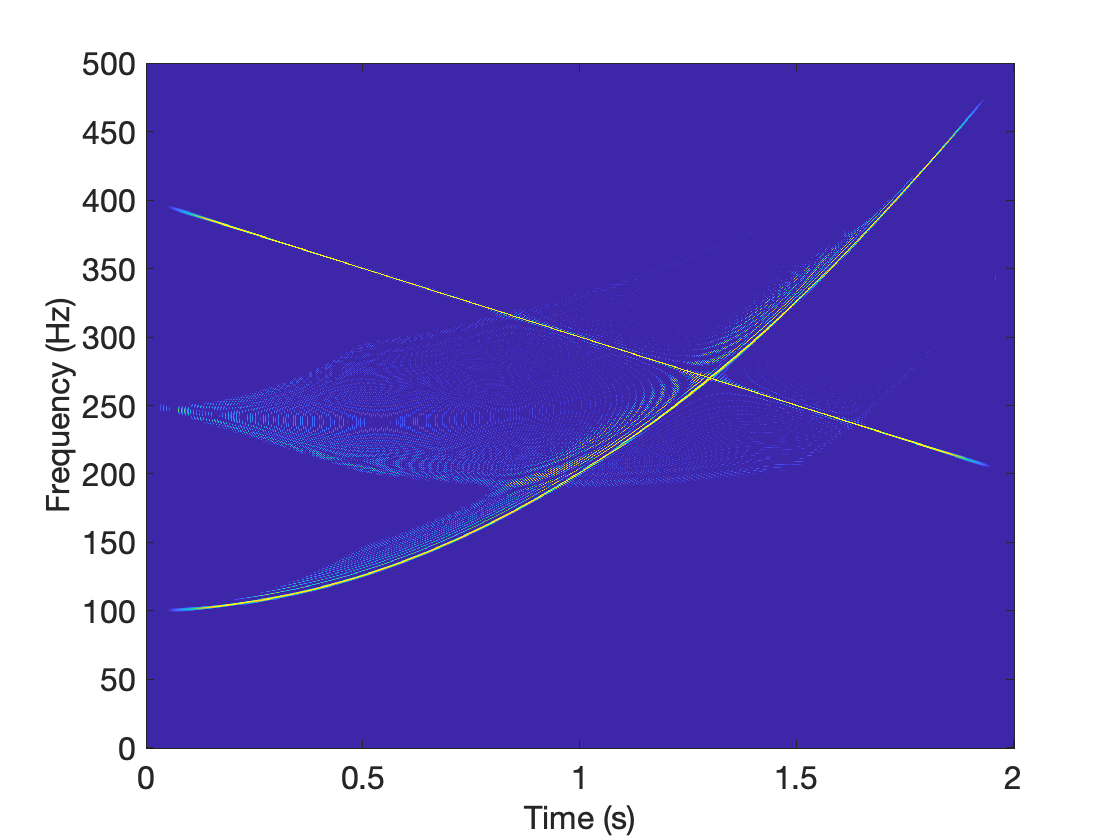}
    \caption{Wigner-Ville distribution of the double chirp signal as in Fig. \ref{fig:spectrogram}. Very visible are both the improved resolution on the chirps and the interference patterns.
    }
    \label{fig:wigner}
\end{figure}

It does look quite different from the previous ones, as it involves a cross-correlation of the time series with its own time-reversed version. Notice that for each $t$ values distant in time have the same weight of values nearby, indicating that this is a highly non-local transformation. An example can be seen in Fig. \ref{fig:wigner}, where the WV transform of the same double chirp as in Fig. \ref{fig:spectrogram} is shown. Very visible are the two important aspects of the distribution: the positive one is that the chirps are sharper in both time and frequency  if compared with the spectrogram, while the negative one is there are spurious ghost patterns in the image. A more serious problem of the ghosts is that the statistics to use for the detection of significant features is not clear, as it can be demonstrated that with the exception of very special cases, the Wigner-Ville distribution cannot be everywhere positive. 

A generalization of the WV distribution was introduced by Cohen \cite{cohen}, who showed that all time-frequency representations can be expressed in the form of the {\it Cohen class} shown in Eqn. \ref{eq:cohen}:

\begin{equation}
C(t,\omega) = {1\over 2\pi^2}\iiint f^*(t-{1\over 2}\tau) f(t+{1\over 2}\tau) \phi(\theta, t) e^{-i\theta t-i\tau\omega+i\theta u} du\,\, d\tau\,\, d\theta
\label{eq:cohen}
\end{equation}

In the Cohen class, the function $\phi(\theta,\tau)$ is called {\it kernel}. The kernel is what determines the distribution and its properties. There are many possibilities (see \cite{cohen}): if $\phi=1$ we recover the WV distribution, while if 

\begin{equation}
\phi(\theta,\tau) = \int h^*(t-{1\over 2}\tau) h(t+{1\over 2}\tau) e^{-i\theta u} du
\label{eq:cohen_FT}
\end{equation}

we obtain

\begin{equation}
S(t,\omega) = \left| {1\over \sqrt{2\pi}}\int e^{-i\tau\omega} f(\tau) h(\tau-t) d\tau \right|^2
\label{eq:cohen_FT_2}
\end{equation}

which is the functional form of the short-time Fourier transform.

{\bf Acknowledgements}
The authors thank S. Motta for useful comments.
T.M.B. acknowledges financial contribution from PRIN INAF 2019 n.15.


\begin{thebibliography}{99.}

\bibitem{addison} Addison P S (2002) The Illustrated Wavelet Transform Handbook. IoP, Bristol Philadelphia

\bibitem{bellonihasinger} Belloni T, Hasinger G (1990) An atlas of aperiodic variability in HMXB. A\&A, 230, 103--119

\bibitem{belloni1997} Belloni T, van der Klis M, Lewin W H G, et al. (1997) Energy dependence in the quasi-periodic oscillations and noise of black hole candidates in the very high state. A\& A, 322, 857--867

\bibitem{bpk} Belloni T, Psaltis D, van der Klis M (2002) A unified description of the timing features of accreting X-ray binaries. Ap J, 572:392--406

\bibitem{butz} Butz T (2006) Fourier Transformation for Pedestrians. Springer, Berlin Heidelberg 

\bibitem{cohen} Cohen L (1995) Time-frequency Analysis. Prentice-Hall, Upper Saddle River

\bibitem{cooleytukey} Cooley J W and Tukey J W (1965) An algorithm for the machine calculation of complex Fourier series, Math Comp, 19:297--301

\bibitem{eyerbartholdi} Eyer L and Bartholdi P (1999) Variable stars: Which Nyquist frequency?, A\&AS, 135:1--3

\bibitem{jenkinswatts} Jenkins G M and Watts D G (1968) Spectral Analysis and Its Applications. Holden-Day. San Francisco

\bibitem{Leahy} Leahy D A, Darbro W., Elsner R F, et al. (1983) On searches for pulsed emission with application to four globular cluster X-ray sources: NGC 1851, 6441, 6624, and 6712. Ap J, 266:160--170 

\bibitem{lomb76} Lomb N R (1976) Least-squares frequency analysis of unequally spaced data, Ap\&SS, 39:447--462

\bibitem{mallat} Mallat S (2009) A wavelet tour of signal processing, Third Edition: The Sparse Way. Academic Press, San Diego

\bibitem{nowak} Nowak M (2000) Are there three peaks in the power spectra of GX 339-4 and Cyg X-1? MNRAS, 318, 361

\bibitem{percivalwalden} Percival D B and Walden A T (2000) Wavelet methods for time series analysis. Cambridge Univ. Press, Cambridge

\bibitem{press_etal} Press W H, Teukolsky S A, Vetterling W T and Flannery B P (2007) Numerical Recipes - The Art of Scientific Computing, 3rd edition, Cambridge Univ. Press, Cambridge

\bibitem{scargle82} Scargle J D (1982) Studies in astronomical time series analysis. II. Statistical aspects of spectral analysis of unevenly spaced data, ApJ, 263:835--853

\bibitem{vdk88} van der Klis M (1988) Fourier Techniques in X-ray Timing. In: Timing Neutron Stars, eds. H. Ogelman and E.P.J. van den Heuvel. NATO ASI Series C, Vol. 262, p. 27-70. Kluwer, Dordrecht

\bibitem{vanderplas2018} VanderPlas J T (2018) Understanding the Lomb-Scargle Periodogram, ApJS, 236:16

\bibitem{weaver} Weaver H J (1989), Theory of Discrete and Continuous Fourier Analysis, Wiley, New York

%
%
%
%
%
\end{thebibliography}
\end{document}